\documentclass[letterpaper]{JHEP3}
\pdfoutput=1

\usepackage{amsmath}
\usepackage{amssymb}
\usepackage{amsfonts}
\usepackage{lscape, graphicx}
\usepackage{cite}       
\usepackage{bm}         
\usepackage{verbatim}

\advance\parskip 1.0pt plus 1.0pt minus 2.0pt   
\def\href#1#2{#2}	

\def\coeff#1#2{{\textstyle {\frac {#1}{#2}}}}
\def\half{\coeff 12}

\def\N{{\cal N}}

\def\R{{\mathbb R}}
\def\S{{\mathbb S}}

\def\tr{{\rm tr}}

\def\nf{n_{\rm f}}
\def\Z{{\mathbb Z}}

\def\Dslash{{\rlap{\raise 1pt \hbox{$\>/$}}D}}

\def\lsim{\mathrel {\vcenter {\baselineskip 0pt \kern 0pt
    \hbox{$<$} \kern 0pt \hbox{$\sim$} }}}
\def\gsim{\mathrel {\vcenter {\baselineskip 0pt \kern 0pt
    \hbox{$>$} \kern 0pt \hbox{$\sim$} }}}

\title{Continuity, Deconfinement, and (Super) Yang-Mills Theory }
\author   
    {
    {
    \def\href#1#2{#2}	
    Erich Poppitz,$^{1,}$\footnote{\email{poppitz@physics.utoronto.ca}} ~
    Thomas Sch\" afer,$^{2,}$\footnote{\email{tmschaef@ncsu.edu}}  ~
    Mithat \"Unsal$^{\; 3,}$\footnote{\email{unsal@sfsu.edu}}~
   \\${}^1$Department of Physics, University of Toronto,   
     Toronto, ON M5S 1A7, Canada
   \\${}^2$Department of Physics, North Carolina State University, 
     Raleigh, NC 27695, USA
   \\${}^3$Department of Physics and Astronomy, SFSU, 
     San Francisco, CA 94132, USA\\
        }
    }%

\abstract{
We study the phase diagram of $SU(2)$ Yang-Mills theory with one adjoint 
Weyl fermion on $\mathbf{\R^3 \times \S^1}$ as a function of the fermion 
mass $m$ and the compactification scale $L$. This theory reduces to thermal 
pure gauge theory as $m\to\infty$ and to circle-compactified (non-thermal) supersymmetric 
gluodynamics in the limit $m\to 0$. In the $m$-$L$ plane, there is a line 
of center-symmetry changing phase transitions. In the limit $m\to\infty$, 
this transition takes place at $L_c=1/T_c$, where $T_c$ is the critical 
temperature of the deconfinement transition in pure Yang-Mills theory. 
We show that near $m=0$, the critical compactification scale $L_c$ can 
be computed using semi-classical methods and that the transition is 
of second order. This suggests that the deconfining
phase transition in pure Yang-Mills theory is continuously connected to 
a transition that can be studied at weak coupling. The center-symmetry 
changing phase transition arises from the competition of perturbative 
contributions and monopole-instantons that destabilize the center, and 
topological molecules (neutral bions)  that stabilize the center. The 
contribution of molecules can be computed using supersymmetry in the 
limit $m=0$, and via  the  Bogomolnyi--Zinn-Justin (BZJ) prescription  
in non-supersymmetric gauge theory. Finally, we also give a detailed discussion of an
issue that has not received  proper attention in the context of $N$$=$$1$ theories---the non-cancellation of  nonzero-mode determinants 
around supersymmetric BPS and KK monopole-instanton backgrounds  on $\R^3 \times \S^1$. We explain why the non-cancellation
 is required for consistency with holomorphy and supersymmetry and perform an explicit calculation of the one-loop determinant ratio.

\smallskip

{\small{

 }}}
 
\begin{document}

\maketitle

 \section{Introduction}

 Consider a quantum mechanical system with a potential with  multiple 
degenerate minima. The ground state energy (as well as the energies 
of higher eigenstates) has a weak coupling expansion of the form 
\begin{equation}
E(g) =  E_{\rm pert.} + E_{\rm non pert.}  
  =   E_0 \left[ 1+O(g)\right] 
     + e^{-1/g} \left[1+ O(g)\right] + O( e^{-2/g})\, . 
\label{expand}
\end{equation}
Since  $e^{-1/g}$ has an essential singularity at $g=0$ it is 
impossible to  express this contribution as a perturbative series in $g$, 
and hence this term\footnote{The exponentially small terms may also be
multiplied by additional  negative powers and logarithms of $g$.} is 
intrinsically  non-perturbative.  Some of the most  
interesting phenomena in quantum mechanics---tunneling, the 
absence of spontaneous symmetry breaking, the formation of 
energy-bands  in periodic  potentials---are due to  $ e^{-1/g} $ 
effects. 
Although the leading term and the exponentially
small contributions  in (\ref{expand}) are intertwined in a deep way, 
as typical inaccuracies of perturbation theory are expressed in terms 
of functions with essential  singularities as above,  
there is a sense in which (\ref{expand}) should be seen as a 
double expansion, a perturbative expansion in $g$ and a 
non-perturbative expansion in  $ e^{-1/g}$. 

 In this paper, we will use this double expansion to study the phase 
diagram of an asymptotically free  gauge theory with strong coupling 
scale $\Lambda$ on $\R^3 \times \S^1$. In a theory without fermions 
the compactification scale on the $\S^1$ circle can always be 
given  a thermal interpretation. At small $\S^1$, of size $L 
\ll \Lambda^{-1}$, it is well-known that  such theories are 
amenable to a perturbative  treatment. A less widely appreciated 
fact is that,  if certain conditions are satisfied, such theories 
are also amenable to non-perturbative semi-classical studies. 
Let $ \Omega= P \exp \left[{i \int_{\S^1} A_4 dx_4}\right]$  
denote the gauge holonomy (or Wilson line) in the compact direction, 
which, classically, is a ``flat direction''.  We expect that quantum 
effects will induce a potential for the holonomy $\Omega$ of the form:
\begin{equation}
V( \Omega)  =  V_{\rm pert.} (\Omega)  + V_{\rm non pert.}(\Omega)~,
\end{equation}
where  $V_{\rm pert.}$ is the contribution of the perturbative 
loop-expansion in $g^2$ and  $V_{\rm non pert.}(\Omega)$  is a 
non-perturbative expansion, presumably containing terms of the form 
$e^{-c/g^2}$. The perturbative term $V_{\rm pert.}$ was initially 
computed in \cite{Gross:1980br}, and the calculation was extended to 
higher order in \cite{Belyaev:1989yt,Enqvist:1990ae,KorthalsAltes:1993ca}.
Although  the perturbative potential $V_{\rm pert.} (\Omega)  $ is by 
now part of the standard books of thermal field theory, $V_{\rm non pert.}
(\Omega)$ has not received as much attention.  
 
 The perturbative calculation of the effective potential for the 
Wilson line  in pure $SU(N)$ Yang-Mills theory on  $\R^3 \times  
\S^1$ with small $L=\beta$ gives \cite{Gross:1980br}:
\begin{equation}
 V_{\rm pert.} (\Omega)   = 
  -   \frac{2}{\pi^2 \beta^4} \sum_{n=1}^{\infty}  \frac{1} {n^4}   
         |{\rm tr} \Omega^{n}|^2 (1+ O(g^2)),
 \label{pert}
  \end{equation}
leading to the conclusion that at small $\beta$ the theory is in 
a deconfined phase, with broken center-symmetry $\langle \frac{1}{N}
\tr \Omega \rangle =1$. If one thinks in terms of eigenvalues of 
$\Omega$,  the potential (\ref{pert}) generates an attraction among 
the eigenvalues. In other words, the effective mass-squared for the 
Wilson line is negative.

 Based on numerical simulations on the lattice we know that the 
deconfinement transition in pure Yang-Mills theory takes place 
at a temperature of  order $\Lambda$: $T_d  =a  \Lambda$ where 
$a$ is a pure number of order one. At one-loop order in perturbation 
theory, (\ref{pert}) shows that  the center-symmetry is broken. 
Higher order corrections do not alter this conclusion; there is 
no effect at any order in perturbation theory that competes
with center symmetry breaking. Hence, the phase transition must be 
induced by  $V_{\rm non pert.}(\Omega)$.  Disregarding such 
non-perturbative effects, one would conclude that one cannot 
explore the transition as the temperature is lowered, from the 
deconfined to the confined phase, using weak coupling techniques.

\begin{FIGURE}[ht]{
    \parbox[c]{\textwidth}
        {
        \begin{center}
        \includegraphics[angle=0, scale=0.5]{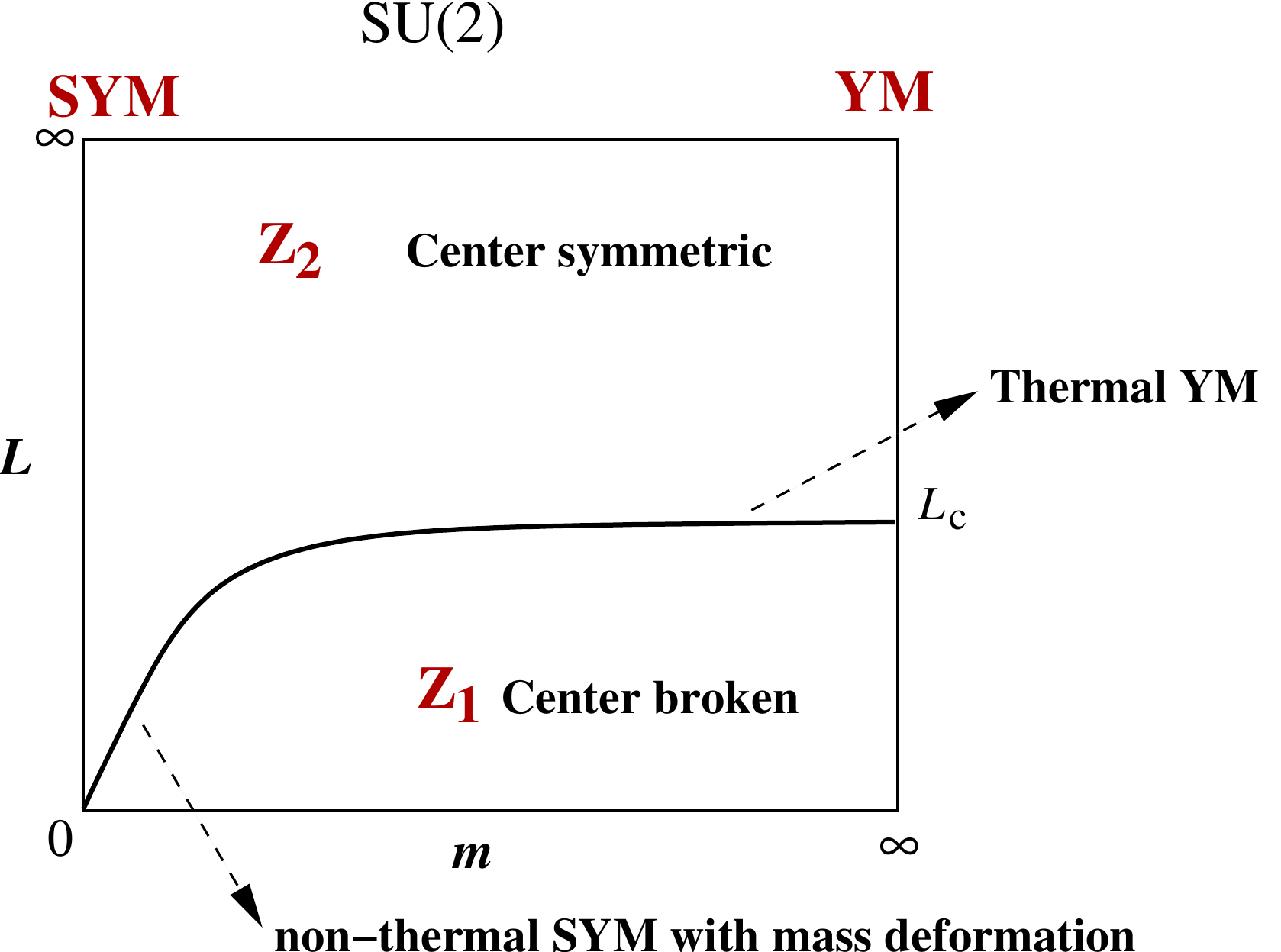}
	\hfil
        \caption
	{The thermal deconfinement phase transition in pure Yang Mills
        (YM) theory can  be accessed through a  non-thermal (quantum) 
        phase transition in supersymmetric Yang Mills (SYM) theory 
        deformed by a gluino mass term. In the massless limit, the 
        supersymmetric theory does not have a phase transition. The 
        phase transition at small-$m$ is analytically calculable and, 
        by decoupling, it is connected to thermal deconfinement phase 
        transition in pure YM theory.} 
        \label{fig:phase2}
        \end{center}
        }
    }
\end{FIGURE}
In this work, we propose a strategy to analytically study the 
center-symmetry changing phase transition in four dimensional gauge 
theories based on an observation discussed in  \cite{Unsal:2010qh}.
The main idea, schematically shown in Figure~\ref{fig:phase2}, is as 
follows: It is well-known that $\N=1$ SYM with periodic boundary 
conditions for fermions does not have a phase transition as a function 
of radius. In fact, for a supersymmetric gauge theory with Hamiltonian 
$H$ and fermion number operator $F$,  
\begin{equation}
\label{Index}
\widetilde Z^{\rm SYM}( L) = \tr \left[ e^{-L H} (-1)^F \right] 
\end{equation}
is the supersymmetric (Witten)  index and is independent of radius. 
In softly broken supersymmetric theory, however, this quantity does 
not have an interpretation as an index.  Consider adding a small  
mass for the fermion  in $\N=1$ SYM. Eqn.~(\ref{Index}) is still 
well-defined, and can be interpreted as a {\it twisted  partition 
function}. The twisted partition function is  a signed sum over the 
states in the bosonic and fermionic  Hilbert spaces, $\cal H_B$ and 
$\cal H_F$, according to the $\Z_2 =(-1)^F$ grading, 
\begin{equation}
\widetilde Z^{\rm SYM}( L,m) =  Z_{\cal B} - Z_{\cal F} 
=  \sum_{n \in \cal H_B} e^{- L E_n}  \; - \; 
 \sum_{n \in \cal H_F} e^{- L E_n}  ~.
 \label{tpf}
\end{equation}
This is different from the ordinary partition function, $Z^{\rm SYM}
(\beta,m) =  Z_{\cal B} + Z_{\cal F}$ by the over-all sign of the 
contribution of  fermionic states.

The twisted partition function, despite being a non-thermal quantity
for general values of  the fermion mass $m$, is immensely useful as
a tool that continuously connects the thermal phase transition in 
pure Yang Mills theory with a semi-classically calculable transition 
on  $\R^3 \times S^1_\beta$. A similar continuity argument at finite 
baryon density was made in \cite{Schafer:1998ef}. For $m \neq 0$,   
(\ref{tpf}) should be viewed as probing the phase structure of the 
theory as a function of radius $L$ (which does not generally have 
an interpretation as inverse temperature). As emphasized,  
the twisted partition function is manifestly non-thermal. Yet,  it can 
be used to study aspects of a genuine  (thermal) deconfinement phase 
transition in certain limits. This is due to the  the following 
decoupling argument.  If the mass of the fermion is infinite, or 
much larger than the  strong scale of $\N=1$ SYM,  $ \widetilde Z 
( L, m) $ reduces to the ordinary thermal partition function of  
pure Yang-Mills theory:
\begin{equation}
\widetilde Z^{\rm SYM} ( L, m) \Big |_{m \rightarrow \infty}  
\Longrightarrow  Z^{\rm YM} ( \beta) = \tr [ e^{-\beta H} ] ~,  
\qquad \beta \equiv L~.
\end{equation} 
In this limit, because the heavy fermion decouples, we  may 
identify the circumference $L$ with the inverse temperature $\beta$.  
For a heavy fermion, the  choice of the boundary condition is immaterial. 

  In this work, we will show that the center-symmetry changing phase 
transition at small $m$ can be computed semi-classically.\footnote{\label{calc}In order to be precise, we note that the small-$m$, small-$L$ calculability  of the transition applies outside of a finite strip around the phase transition line in Figure \ref{fig:phase2}. As usual in second order phase transitions, fluctuations become strong near the critical point and the renormalization group equations describing the critical theory---in the case at hand, in the 3d Ising universality class---are nonperturbative. However, in the weak coupling small-$m$, small-$L$ regime, the width of the strongly fluctuating critical region is small, controlled by powers of the small parameter $L \Lambda$, and the critical values of the mass $m_c$  at fixed $L$ (or  critical size $L_c$  at fixed $m$) can be reliably determined up to small corrections.} In this 
limit the transition takes place at small $L$, as shown in 
Figure~\ref{fig:phase2}. The physics of the transition is quite 
interesting. It is based on the competition between topological 
molecules, called  ``neutral bions" or ``center-stabilizing bions'', 
and semi-classical monopole-instanton effects, as well as perturbative
effects. We will argue that these effects are also present at large 
$m$, in the pure gauge theory, but that in this limit the effect cannot 
be reliably computed using semi-classical methods.

\section{Mass deformation of $\mathbf{\N=1}$ super-Yang-Mills on  
$\mathbf{\S^1 \times \R^3}$}
\label{sec:massatsmallL}

\subsection{Perturbation theory}
\label{sec:perttheory}
 
  Classical vacua of the theory on $\mathbf{\R^3 \times \S^1}$ are 
labeled by the expectation value of the Wilson line
\begin{equation}
 \label{omegadef}
 \Omega = \exp\left[i \int A_4 dx_4\right]~.
\end{equation} 
When $ L \Lambda \ll 1$, non-zero frequency Kaluza-Klein modes are 
weakly coupled and may be integrated out perturbatively. If we 
consider periodic boundary conditions for both the gauge fields
and the adjoint Weyl fermions, $A_\mu(L)=A_\mu(0)$ and $\lambda(L) 
= + \lambda(0)$, the one-loop effective potential for the Wilson
line is \cite{Meisinger:2001fi,Unsal:2010qh}:
\begin{align}
V^{\rm SYM}_{\rm pert.}[\Omega,  m ] &= \frac{2}{\pi^2L^4}
    \sum_{n=1}^{\infty} \>
	\left[ - 1 + \frac{1}{2} \,  (nLm)^2 K_2(n L m) \right]
	\frac{\left|\tr\, \Omega^n \right|^2}{n^4} \, .
\label{eq:pot1}
\end{align}
Here $m$ is the fermion mass and $K_2(z)$ is the modified Bessel 
function of the second kind, with asymptotic behavior
\begin{eqnarray}
    K_2(z) =
    \left\{ \begin{array}{ll}
     \frac{2}{z^2} - \half + O(z^2)\,, \quad &  z \ll 1 \,;
     \\
     \sqrt \frac{\pi}{2z} \; e^{-z}\,, \quad & z \gg 1 \,.
    \end{array} \right.
\label{asymptotes}
\end{eqnarray}
As the mass $m \to \infty$, the fermions decouple regardless of their 
boundary conditions, and the effective potential (\ref{eq:pot1}) reduces 
to the pure gauge result given in (\ref{pert}), with the identification 
$L=\beta$: 
\begin{eqnarray}
    V^{\rm SYM}_{\rm pert.}[\Omega,  m] \big |_{m \rightarrow \infty} = 
    V^{\rm YM}_{\rm pert.} [\Omega] (1+ O(e^{-Lm}))~.
\end{eqnarray}
In the opposite limit of massless fermion, the one-loop potential vanishes:
\begin{eqnarray}
     V^{\rm SYM}_{\rm pert.}[\Omega,  m=0] =0~.
\end{eqnarray}
In fact, because of supersymmetry and the fact that (perturbatively) 
the theory possesses a moduli space of vacua, the $m=0$ SYM theory does 
not generate a potential for the Wilson line to any order in perturbation 
theory. At a typical point on the moduli space, a nonzero background  
Wilson line is turned on:  
\begin{eqnarray}
\Omega= \left( \begin{array}{cc} e^{i \Delta\theta/2} & \\
&  e^{- i  \Delta \theta/2} 
\end{array} \right)~,
\label{bcgr0}
\end{eqnarray}
and  the $SU(2)$ gauge group abelianizes down to $U(1)$. Here, $\Delta 
\theta$ is the separation between the eigenvalues of the Wilson line. 
Non-perturbatively, the moduli space is lifted due to effects leading 
to $\Delta \theta =\pi$, i.e., to a center-symmetric holonomy. This effect 
will be described in two complementary ways in the  next Section. 

 We now  turn on a small mass corresponding to soft supersymmetry 
breaking, $m \ll \Lambda$. In the small $L \Lambda \lesssim 1$ regime, 
we also have $m L \ll 1$.  In this case,  using the  small-$z$ asymptote 
of (\ref{asymptotes}), we observe that the leading term at  $O(m^0)$ 
cancels and an $O(m^2)$ potential is induced. The effective potential 
becomes (up to $O(m^4)$ corrections):
\begin{eqnarray}
  V_{\rm SYM}[\Omega]= -
        \frac{m^2}{2 \pi^2L^2}
        \sum_{n=1}^{\infty}   \frac{1}{n^2} \,
        \left|\tr\, \Omega^n \right|^2   
      =  -  \frac{m^2}{L^2} B_2 \left( \frac{ \Delta  \theta}{2 \pi} \right) ~,
\label{eq:pot4}
\end{eqnarray}
where $B_2(x)= x^2 -x + \frac{1}{6}$ is the second Bernoulli polynomial (the 
last equality above is valid when $\Delta \theta\in [0,2\pi]$).  

Within the domain of validity of the perturbative analysis, (\ref{eq:pot4})  
shows that Wilson lines with all winding numbers are unstable when the 
fermion mass is non-zero, despite the use of periodic boundary condition 
for fermions. Consequently, the $\nf = 1$ theory  at any non-zero mass $m$
and sufficiently small $L$ will have completely broken center symmetry. 
On the other hand, the fact that the one loop potential is small, of $O(m^2)$,
implies that exponentially small semi-classical effects can compete with the 
perturbative potential.

\subsection{Non-perturbative effects at $\mathbf{m=0}$ via supersymmetry} 
\label{susy}

 In the classical background (\ref{bcgr0}), and at weak coupling, the  
Wilson line (\ref{omegadef}) behaves as an adjoint Higgs field.  The 
theory at short distances is described by non-abelian $SU(2)$, and at 
long distances, it is described by an abelian $U(1)$ subgroup. Using 
abelian duality, $ \epsilon_{ \mu\nu \lambda} \partial_{\lambda} \sigma  
= {4 \pi L \over g^2 } F_{\mu\nu}$ for $\mu,\nu=1,2,3$, we may map the 
gauge field to a  spin-zero dual-photon $\sigma$. It is also useful 
to define  the exponent of gauge holonomy (\ref{bcgr0}):
\begin{equation}
 \label{defofb}
  b \equiv \frac{4 \pi}{g^2} \Delta \theta~.
\end{equation} 
The  kinetic terms of the fields $\sigma$ and $b$ are 
\begin{equation}
{\cal L} = {1 \over 2} {g^2\over (4 \pi)^2 L}
 \left[(\partial_i b)^2 + (\partial_i \sigma)^2\right]\, . 
\end{equation} 
In terms of superfields this corresponds to a K\" ahler potential:
\begin{equation}
\label{kahler}
  K = {g^2\over 2 (4 \pi)^2 L}\; {\mathbf{B^\dagger B}}~,
\end {equation} 
for the chiral superfield $\mathbf{B}$, whose lowest component\footnote{We note that the relation between $\bf{B}$ and $b$ is, in fact, nonlinear, see eqn.~(\ref{x2}), due to the perturbative corrections to the moduli space metric along the Coulomb branch. These are also reflected in  the non-cancellation of the one-loop fermion and boson nonzero mode determinants around the BPS and KK monopole-instantons, see Appendix \ref{nonzerodets} for a detailed discussion. As these subtleties represent subleading corrections to the K\" ahler metric (\ref{kahler}), we ignore them in the main text.} 
 is 
$b - i \sigma$; the fermionic component is the component of the gluino 
field $\lambda$ which remains massless along the Coulomb branch 
(\ref{bcgr0}). The effective Lagrangian following from (\ref{kahler}) 
gives the long-distance perturbative description of the theory on $\R^3 
\times \S^1$ -- essentially a massless free-field theory.
 
 The non-perturbative dynamics of the theory is quite rich: due to 
the compact topology of the ``adjoint Higgs" (\ref{omegadef}), there 
are two types of elementary monopole-instantons,  ${\cal M}_{\rm 1}$ 
and  ${\cal M}_{2}$. These are sometimes called 3d instanton and 
twisted-instanton, or BPS-monopole-instanton and KK-monopole-instanton, 
see~\cite{Lee:1997vp, Kraan:1998sn}. The Nye-Singer index theorem implies 
two fermionic zero-modes for each \cite{Nye:2000eg,Poppitz:2008hr}. The 
4d BPST instanton (in the long distance regime) can be viewed as a 
composite of these two. The monopole-instantons give rise to 't Hooft 
vertices, or amplitudes, of the form:
\begin{eqnarray}
&&{\cal M}_{1} =   e^{ - b +i \sigma}     \lambda \lambda, \qquad 
  {\cal M}_{2} = \eta  e^{ + b -i \sigma} \lambda \lambda,  \cr  \cr
&&\overline {\cal M}_{1} =  e^{ - b - i \sigma} 
                             \bar \lambda \bar \lambda, \qquad 
  \overline {\cal M}_{2} = \eta  e^{ + b + i \sigma}
                             \bar \lambda \bar \lambda, 
\label{1amplitudes}
\end{eqnarray}
where  $\eta=  e^{- \frac{8\pi^2}{g^2}} = e^{-2S_0}$ is the 4d instanton 
amplitude (we set the topological theta angle to zero). Since 
(\ref{1amplitudes}) carry just two zero modes, they generate a 
superpotential, given by:
 \begin{equation}
 W_{\R^3 \times \S^1}  = {M_{PV}^3 L \over g^2} 
     \left( e^{- {\bf B}} +  \eta e^{\bf B} \right) 
  =  {2 M_{PV}^3 L \over g^2}\;  e^{-S_0} \cosh 
    \left( {\bf B} - \frac{4 \pi^2}{g^2} \right)~,
\label{2vac1}
\end{equation} 
where the coupling is normalized at the cutoff scale $M_{PV}$; details 
of the instanton calculation leading to (\ref{2vac1}) can be found 
in \cite{Davies:2000nw}. Thus, the infrared Lagrangian is given by 
(\ref{kahler}) and (\ref{2vac1}):  
\begin{eqnarray}
\label{lagr1}
{\cal L} =  \int d^4 \theta  \; K   
   +  \left( \int d^2 \theta \;  W +  h.c.\right)~,
\end{eqnarray}
and the scalar potential can be easily found:
\begin{eqnarray}
\label{pot1}
 V (b, \sigma) \equiv 
  K_{\mathbf{B^\dagger B}}^{-1}  \left|\frac{\partial W}{\partial  {\bf B}} \right|^2  
  =  {64 \pi^2 M_{PV}^6 L^3  e^{-2S_0}  \over g^6} 
   \left(   \cosh \left( 2\left(b - \frac{4 \pi^2}{g^2}\right) \right) 
     -    \cos 2 \sigma \right),
\label{bosonic}
\end{eqnarray}
where $K_{\mathbf{B^\dagger B}}$ is the mixed second derivative of the K\" ahler potential.
Furthermore, it is convenient to introduce, instead of $b$ of 
eqn.~(\ref{defofb}), the shifted field:
\begin{equation}
\label{defofbprime}
b \equiv {4 \pi^2 \over g^2} + b^\prime~, \hspace{0.5cm} 
{\Delta \theta \over 2\pi} = {g^2 \over 8 \pi^2}\; b^\prime + {1 \over 2}~.
\end{equation}
Finally, to rewrite the potential in terms of the strong-coupling scale 
$\Lambda$ and the $\S^1$ size $L$, we use the relations:
\begin{equation}
\label{rgrelations}
{M_{PV}^3 \over g^2} e^{- {4 \pi^2 \over g^2}} = \Lambda^3~,\hspace{0.5cm}
{4 \pi^2 \over g^2}  \approx  3 \log {1 \over \Lambda L}~. 
\end{equation}
These relations express the fact that the scale for the coupling  
of the effective theory is set by the compactification scale $L$.
We use the one-loop running coupling constant in the prefactors   
and two-loop running in the exponent of the instanton amplitude.

Thus, using (\ref{defofbprime}) and (\ref{rgrelations}), we obtain the 
final expression for the scalar potential (\ref{pot1}): 
\begin{equation}
\label{potentialbion}
V_{\rm bion} (b^\prime, \sigma) = 
  48  L^3 \Lambda^6 \log {1 \over \Lambda L}  
    \left( \cosh 2 b^\prime - \cos 2 \sigma \right) ~.
\end{equation}
Note that if the superpotential $W$ is determined by BPS and KK 
monopole-instantons then the scalar potential is governed by (correlated) 
monopole-anti-monopole pairs. We have therefore denoted the scalar 
potential by $V_{\rm bion}$. We will make this relationship more 
explicit in the following Section.

 The potential $V_{\rm bion}$ for the $b^\prime$-field, $\sim\cosh(2b^\prime)$, 
generates a non-perturbative repulsive interaction between the eigenvalues 
of the Wilson line around $\S^1$: it is minimized at $\langle b^\prime \rangle 
= 0$, which, from (\ref{defofbprime}), corresponds to maximally separated 
eigenvalues, $\langle \Delta \theta\rangle = \pi$, or in terms of the 
Wilson line:
\begin{equation}
\langle \Omega \rangle = \langle e^{i A_4 L} \rangle = 
 \left( \begin{array}{cc}   
  e^{i \frac{\pi}{2}} &  \\ & e^{-i \frac{\pi}{2}}  
 \end{array}    \right) , \qquad 
 \langle \tr \Omega \rangle =0~,
  \label{minimum} 
\end{equation} 
up to gauge rotations. This is the center-symmetric vacuum of the theory 
on $\R^3 \times \S^1$. On the other hand, the $V_{\rm bion}$ potential for 
the $\sigma$ field has two minima, located at 
\begin{equation}
\langle \sigma \rangle =\{0, \pi\} ~,
\end{equation}
associated with discrete chiral symmetry breaking. Evidently, in the 
effective lagrangian, the mass gap for gauge fluctuations  $\sigma$ 
is generated by the operator $e^{ \pm  2i \sigma }$, and for the spin-zero 
scalar (the fluctuation of gauge holonomy $b'$) it is generated by   
$e^{ \pm  2b'}$.  

 Expanding the action around the center-symmetric gauge holonomy and 
using component notation, we find the effective lagrangian:
\begin{eqnarray}
\label{lagr2}
{\cal L}=&&  {1 \over 2} {g(L)^2 \over (4 \pi)^2 L}
  \left[(\partial_i b')^2 + (\partial_i \sigma)^2\right] 
 +  i  \frac{L}{g^2} \bar \lambda \sigma_i \partial_i \lambda  
 +    \alpha  e^{- {4 \pi^2 \over g^2(L)}} \left[ 
   \left( e^{-b'+i \sigma} +  e^{+b'-i \sigma}\right ) \lambda \lambda  
 +  {\rm c.c.}  \right] 
 \cr \cr
&& \mbox{}+ \beta \; \frac{e^{-   {8 \pi^2 \over g^2(L)}}}{L^3}  
   \left[ e^{- 2b'} +  e^{ 2b'} -   e^{- 2i \sigma }  -  e^{ 2i \sigma } 
\right]   ~.
\end{eqnarray}
We expressed  the Lagrangian in  component notation to elucidate the  
physical origin of the various terms\footnote{In order not to clutter 
notation, we kept only the exponential dependence of the coupling 
$g^2(L)$ in the non-perturbative terms in (\ref{lagr2}), i.e., we absorbed  
the numerical coefficients and  the $\log {1\over L \Lambda}$ dependence 
in the prefactors $\alpha$ and $\beta$; these can be recovered from 
(\ref{2vac1}), (\ref{potentialbion}).}.

\subsection{Non-perturbative effects at $\mathbf{m=0}$ via 
topological molecules}
\label{susywithoutsusy}

  We now provide  a derivation of the bosonic 
potential without the use of supersymmetry. This formalism will apply to 
both supersymmetric and non-supersymmetric theories. As discussed in 
Refs.~\cite{Unsal:2007vu,Unsal:2007jx}, in general gauge theories with 
massless adjoint fermions (i.e., not only in  the supersymmetric single 
massless Weyl flavor case $n_{\rm f} = 1$), the 3d instanton and twisted 
instanton do not  generate a mass gap for the gauge fluctuations because 
of  their fermionic zero mode structure. The zero modes are determined 
by the  index theorem \cite{Nye:2000eg,Poppitz:2008hr}, and the corresponding 
instanton amplitudes have the form given in (\ref{1amplitudes}) or the 
generalization thereof for the $n_{\rm f} > 1$ case. 

 Unlike the superpotential, which arises due to monopole-instantons with 
exactly two zero modes, the associated bosonic potential must be induced 
by topological molecules which do not have any fermionic zero modes. The 
bosonic potential is generated  by correlated monopole-anti-monopole pairs. 
To second order in the semi-classical expansion, the possibilities are the 
following: $[{\cal M}_{1} \overline {\cal M}_{1}]$, $[{\cal M}_{2}
\overline{\cal M}_{2}]$, $[{\cal M}_{1}\overline{\cal M}_{2}]$, $[{\cal M}_{2} 
\overline{\cal M}_{1}]$. These objects can be viewed as composites of  
(\ref{1amplitudes}). The magnetic and topological  charges  and the 
amplitudes associated with these instanton-anti-instanton events are: 
\begin{eqnarray}
{\rm composite}  \qquad &  \left(Q_m,  Q_T \right) \qquad  &   {\rm amplitude}
   \qquad     \cr   \cr
[{\cal M}_{1} \overline {\cal M}_{1} ]  \qquad  &   (0, 0)  \qquad   
    & e^{-2b'}  \qquad      \cr \cr 
[{\cal M}_{2} \overline {\cal M}_{2} ]  \qquad  &  (0, 0)   \qquad   
    &  e^{+2b'} \qquad     \cr  \cr 
[{\cal M}_{1} \overline {\cal M}_{2} ]  \qquad  &  (+2, 0 ) \qquad   
    & e^{+2i \sigma} \qquad     \cr \cr
[ {\cal M}_{2} \overline {\cal M}_{1} ] \qquad  & (-2, 0)  \qquad   
    & e^{-2i \sigma}   \qquad ~.
\label{composites}
\end{eqnarray}
The action and interaction due to massless boson exchange of two 
fundamental monopole instantons with (bosonic) amplitudes 
$e^{n_b b'+ i n_m \sigma}$ and $e^{n'_b b'+ i n'_m \sigma}$ is (here $n_b,n_m 
= \pm 1$):
\begin{equation}
S(r)= 2S_0  + S_{\rm  int} =  \frac{8 \pi^2}{g^2} 
    +  \left(-n_b n_b'+ n_m  n_m'\right) \frac{4 \pi L}{g^2 r}
\label{qzm1}
\end{equation}
where $r= |{\bf r}|$ is the separation between two instanton events. 
The interaction proportional to $n_b n_b'$ is due to the exchange 
of the massless $A_4$ modulus and the one proportional to $n_m n_m'$
is due to the exchange of the dual photon (``magneto static"). The 
interaction term is repulsive for $(-n_b n_b'+ n_m  n_m') > 0$, 
attractive for  $(-n_b n_b'+ n_m  n_m') <0$ and zero otherwise. The 
fermion-induced interactions will be considered below.

\subsubsection{Magnetic bions, quasi-zero modes, and the mass gap 
for the dual photon} 

 All the topological molecules contributing to the bosonic potential 
have vanishing topological charge,  i.e., they are indistinguishable 
from the perturbative vacuum in that sense.  However, the $[{\cal M}_{1} 
\overline {\cal M}_{2} ]$ (and its anti-molecule) events carry two units 
of magnetic charge. The prefactor of the amplitude can be found as follows. 
 
 The ${\cal M}_{1}$ instanton has four bosonic zero modes. Three of these 
are the positions ${\bf x} \in \R^3$ and one is related to the internal 
$U(1)$ symmetry. Note that there is no size modulus associated with 
monopole-instantons, unlike the 4d BPST instanton. This is one of the 
reasons that we can do reliable semi-classical analysis.  Let an 
$\overline {\cal M}_{2} $ be located at ${\bf y}$. Because of the 
interaction which depends on the separation  of the two events, 
${\bf r=x-y}$, the relative coordinate is no longer an exact zero mode,
while  the ``center  of mass" position  ${\bf R =(x+y)}/2$ of the two 
events is an exact zero mode. The relative coordinate is now a quasi-zero 
mode. This is to say that the operator of the quadratic fluctuations in 
the background of  ${\cal M}_{1}$ and  $\overline {\cal M}_{2} $  has, in 
its spectrum, an exact zero mode, a low lying quasi-zero mode and 
parametrically separated Gaussian fluctuations.  The latter modes can 
be trivially integrated out. The zero and quasi-zero modes are particularly 
important. In particular, the integrals over the quasi-zero modes need to 
be done exactly. 
  
 The magnetic bion amplitude associated with a bion located at ${\bf R}$ 
can be found by integrating over the quasi-zero mode exactly. Below, we  
write the expression for the $n_{\rm f}$ flavor theory for later convenience. 
The amplitude associated with an $[{\cal M}_{1} \overline {\cal M}_{2} ]$ 
composite is:
\begin{eqnarray} 
 [{\cal M}_{1} \overline {\cal M}_{2} ] \sim 
 {\cal A} e^{-2S_0} e^{2i \sigma},  
\end{eqnarray}
where
\begin{eqnarray} 
\label{lambda1}
 {\cal A} =   \int d^3r\,  
     e^{- \left(  2 \times \frac{4\pi L}{g^2  r} + 4n_f \log r \right)   }   
=  4 \pi I (\lambda, n_f)~,\qquad  \lambda \equiv {g^2 \over 8 \pi L}~. 
\end{eqnarray}
The meaning of the terms in the exponent is as follows: $2 \times 
\frac{4\pi L}{g^2  r}$ accounts for the repulsion due to exchange of 
$\sigma$ and $b$-scalars, and $4n_f \log r$ is the attraction due to 
fermion zero mode exchange. Consequently, there is a single saddle-point 
in the  quasi-zero mode integral, given (for $n_{\rm f} = 1$) by:
\begin{equation}
r_{\rm b} = \frac{4 \pi L}{g^2}~,
\label{bionsize}
\end{equation}
which can be interpreted as the magnetic bion size. The bion size
is much larger than monopole-size, but much smaller than (uncorrelated) 
inter-monopole separation. Consequently, a representation of the partition 
function as a dilute gas of magnetic bions is justified. The integral in (\ref{lambda1}) is 
given by: 
\begin{eqnarray} 
I(\lambda, n_f) = \int dr  \,      
e^{- \left( \frac{1}{ \lambda r}  + (4n_f-2) \log r  \right)   }  
= \lambda^{4 n_f -3} \Gamma(4n_f-3), 
\label{m-bion}
\end{eqnarray}
where $n_{\rm f} = 1$ for SYM. The  way to check that this is self-consistent 
is as follows. The interaction term for the magnetic bion  in (\ref{qzm1}) 
must be  parametrically smaller than the  leading action in order for the 
relative coordinate to deserve the name quasi-zero mode. Just saying that 
$2S_0 \gg S_{\rm int}$, or equivalently, $r \gg \frac{L}{\pi}$, is not 
sufficiently good, because this does not preclude the $O(1)$ changes 
in the combined action, $2S_0$. It must be such that $|S_{\rm int}| \sim 
O(g^2) \times (2S_0) \ll (2S_0)$. Indeed, the action of the magnetic bion 
configuration associated with $r=r_{\rm b}$, given in (\ref{qzm1}), takes 
the form
\begin{equation}
S(r_{\rm b})= 2S_0 \left( 1 + O(g^2) \right)
\label{qzm2}
\end{equation}
Since the interaction  changes the action only by parametrically 
small $O(g^2)$ effects, the magnetic bion  topological molecule can be 
viewed as a quasi-solution. 
 
The magnetic bion molecules described in this Section are responsible 
for the generation of a mass gap for the dual photon (the potential 
for $\sigma$ in (\ref{lagr2})) in SYM on $\R^3 \times \S^1$, and they
generate the confining string tension. In the following Section we 
will study a second type of topological molecule, which is more subtle 
to identify, but plays an important role in the center-symmetry 
realization. 
 
\subsubsection{Neutral (center-stabilizing) bions and the BZJ prescription}
\label{sec:centersymmetry}

Consider now the other possible composite from the list (\ref{composites}),  
the $[{\cal M}_{1} \overline {\cal M}_{1} ]$ composite which carries no 
magnetic and topological charge (the $[{\cal M}_{2} \overline {\cal M}_{2}]$ 
is treated similarly). Here, the integral over the quasi-zero mode is, 
naively:
\begin{eqnarray} 
[{\cal M}_{1} \overline {\cal M}_{1} ] \sim 
 {\cal A}_{\rm naive} e^{-2S_0} e^{\pm 2b' } , 
\end{eqnarray}
where:
\begin{eqnarray}
 {\cal A}_{\rm naive}(g^2) &=&   \int d^3r   \;        
 e^{- \left(  - 2 \times \frac{4\pi}{g^2  r}  + 4n_f \log r \right)   }   
=  4 \pi \tilde I(\lambda, n_f) ~,\qquad  \cr
\tilde I(\lambda, n_f) &=&  \int dr  \,      
e^{- \left( - \frac{1}{ \lambda r}  + (4n_f-2) \log r  \right)   }\label{e-bion}~.
\label{naive}
\end{eqnarray}
Now, the interactions  between constituents due to $\sigma$ and $b$ 
exchange are both attractive, while the fermion zero mode induced 
attraction is not altered (it remains attractive). The integral, 
(\ref{naive}), as it stands, is dominated by the small $r$ regime, 
where not only (\ref{e-bion}) is incorrect, it is also hard to make 
sense of constituents as the interaction becomes large. This is in 
sharp contrast with the magnetic bion (\ref{m-bion})  
\cite{Unsal:2007jx,Anber:2011de}.  

 At first sight, this may seem to prevent us from computing the 
contribution from these pairs, but this is not actually the case. 
We will reach a satisfactory resolution of the problem, via the 
{\it Bogomolnyi--Zinn-Justin  prescription:} The integrals over 
the quasi-zero modes of attractive instanton--anti-instanton 
molecules can be calculated in a manner initially described 
by Bogomolnyi \cite{Bogomolny:1980ur} in the context of quantum 
mechanics. The relation between this prescription and the Borel 
procedure was pointed out in the same context by Zinn-Justin 
\cite{ZinnJustin:1981dx}, see also \cite{Schafer:1996wv}. The 
prescription is to analytically continue the coupling $g^2$ in the 
instanton-anti-instanton interaction to $-g^2$. This turns the 
attractive Coulomb force into a repulsive one. Then we calculate the 
resulting integral exactly, without any Gaussian approximations. Finally, 
we analytically continue the final result back to positive $g^2$. 

 Recall that, very often for a non-Borel summable series, when $g^2$ 
is continued to $-g^2$, the series become Borel summable. However, 
one needs to continue back to positive $g^2$. Depending on the path 
that one takes the coupling to the positive $g^2$, the Borel sum  
typically produces an ambiguous (non-perturbative) imaginary part.  
This is a manifestation of non-Borel summability.  In the quantum 
mechanical  examples that Refs.  \cite{Bogomolny:1980ur} and 
 \cite{ZinnJustin:1981dx} studied, the ambiguity in the Borel sum 
is canceled by the ambiguity associated with the  attractive 
instanton-anti-instanton molecule. The prescription for the topological 
molecules may be viewed as consistently extending the Borel prescription 
for perturbative sums to non-perturbative sectors with vanishing quantum 
numbers.  For a fuller discussion of these phenomena in field theory,  
see \cite{Argyres:2012}.

  This prescription will give an overall phase between the magnetic 
bion amplitude and center-stabilizing bion amplitude. This phase 
difference is physical and crucial for our considerations.  Following   
the prescription, when we modify $\tilde I(\lambda, n_{f}) \rightarrow 
\tilde I (-\lambda, n_{f})$ (recall that $\lambda \equiv g^2/8 \pi L$) 
the Coulomb-interaction becomes repulsive and we can evaluate the 
integral over the quasi-zero mode. In fact, it is equal to the integral 
$I(\lambda, n_{f})$ for the magnetic bion computed in the previous 
Section. Next, we substitute $g^2\rightarrow -g^2 $ ($\lambda \rightarrow 
- \lambda$) to obtain the final result for the center-stabilizing bion 
amplitude. To  summarize, the  generalization of the BZJ prescription 
to field theory  results in the chain:
\begin{align}
\tilde I (\lambda, n_{f}) \;\; \rightarrow \;\; 
\tilde I (-\lambda, n_{f}) =  I(\lambda, n_{f}) \;\; \rightarrow \;\;  
I(-\lambda, n_{f}) = \left(- \lambda  \right)^{3-4n_f} \Gamma (4n_f-3) 
= - I(\lambda, n_{f}) \, . 
\label{sign}
\end{align}  
The last equality is only valid for integer $n_f$ and gives an overall 
sign of the center-stabilizing bion amplitude opposite that for the 
magnetic bion. We note that  this line of reasoning has a close parallel 
in supersymmetric quantum mechanics \cite{Balitsky:1985in}.
 
 The importance of the relative sign between the magnetic bion amplitude 
and center-stabilizing bion amplitude is worth noting, as it is a physical 
consequence of our prescription. As a result, we obtain for their combined 
contribution:
\begin{eqnarray}
V ( b, \sigma)& \sim & \eta \cosh 2 b' - \eta \cos 2 \sigma 
  = e^{-2S_0} \left[ (1+ 2b'^{\,2}  + \ldots) - (1- 2\sigma^2  + \ldots) \right]
  \nonumber  \\[0.2cm]
 &  =&  2 e^{-2S_0}  \left( b'^{\,2}  +  \sigma^2 \right) ~ ,
  \label{bosonic3}
\end{eqnarray}
the same result that we obtained earlier by using holomorphy. The 
crucial point here is the cancellation of the ``cosmological constant"  
term in the potential. Recall that in a supersymmetric theory with 
unbroken supersymmetry, the expectation value of Hamiltonian is 
positive semi-definite and  $\langle \Psi_n|H | \Psi_n\rangle \geq 0$, 
and that the bound is saturated for the ground state  $\langle \Psi_0|
H | \Psi_0\rangle = 0$.  If the relative sign was not present,  
the ground state energy would not vanish, implying a breakdown of 
supersymmetry. Equally importantly,  the absence of the relative 
sign between the  $[{\cal M}_{1} \overline {\cal M}_{1} ]$ and  
$[{\cal M}_{1} \overline {\cal M}_{2} ]$ amplitudes would lead to 
the presence of a relative sign between the mass term for the two 
scalars of the form $ \left( b'^{\,2}- \sigma^2\right)$, signaling 
an instability.  Clearly, neither is the case. 

\subsection{Center-stabilizing vs. center-breaking  effects in softly 
broken SYM}

 As explained in Section \ref{sec:perttheory} there is no perturbative 
contribution to the Wilson line effective potential for $m=0$, but 
there is a non-perturbatively induced  potential which ensures 
unbroken center symmetry in the supersymmetric theory on $\R^3 \times 
S^1$. This potential, as explained above, is due to center-stabilizing 
bions.

 We now turn on a small but non-zero $m$. At small $L$, one expects a  
competition between the one-loop $O(m^2)$ potential for the Wilson line 
and the non-perturbatively induced superpotential, leading to non-uniformity 
in the $m\to 0$ and $L \to 0$ limits. Taking $m\rightarrow 0$ first, the 
theory lands on the center-symmetric phase. If, instead, the  $L\rightarrow 
0$  limit is taken first, the theory lands on the  center-broken phase. 
The transition line separating center-symmetric and center-broken phases  
must emerge from the $L=m=0$ corner of the phase diagram, as illustrated 
for an $SU(2)$ theory on  Fig.~\ref{fig:phase2}. Let us now describe the 
center-symmetry breaking dynamics in some more detail. 
  
 Adding a soft mass term for the fermions reduces the $\N=1$ supersymmetry  
to $\N=0$ and  has the effect of lifting the fermion  zero modes from the 
instanton amplitudes (\ref{1amplitudes}). The mass perturbation is:
\begin{equation}
\label{lm}
 \Delta {\cal L}_m =   {m \over g^2} \; \tr \lambda  \lambda  + {\rm h.c.}~.
\end{equation} 
The insertion of mass terms lifts the zero modes of the monopole-instanton 
amplitudes (\ref{1amplitudes}) which now contribute to the potential for 
$\sigma$ and $b$. The corresponding calculation is presented in the 
appendix, and  the  result for the monopole-instanton contribution 
to the scalar potential, to leading order in $m$ ($m L \ll 1$), takes 
the form:
\begin{equation}
\label{monopolepotential}
 V_{\rm mon.} = 24 m  L \Lambda^3 \cos\sigma 
  \left( \log {1 \over \Lambda L} \cosh b^\prime 
   - {1 \over 3} b^\prime \sinh b^\prime \right)~.
\end{equation}
Despite the addition of a fermion mass term, the fermion-attraction 
mechanism giving rise to magnetic and center-stabilizing bions is still 
operative, provided the fermion mass is smaller than the inverse size of 
the bions, $r_{\rm b} = {4 \pi L \over g^2}$ from (\ref{bionsize}), i.e., 
for $m L  < {g^2 \over 4 \pi}$ (below, we shall see that this condition 
is obeyed in the regime where we can study the competition between 
center-breaking and center-stabilizing effects). Thus, adding the bion 
and monopole non-perturbative contributions (eqns.~(\ref{potentialbion}) 
and (\ref{monopolepotential}), respectively) to the  the perturbative 
contribution (\ref{eq:pot4}), recalling (\ref{defofbprime}), we obtain 
the full scalar potential: 
\begin{eqnarray}
 \label{totalpotential}
 V_{\rm total} &=& 
  48  L^3 \Lambda^6 \log {1 \over \Lambda L}  
          \left( \cosh 2 b^\prime - \cos 2 \sigma \right)  \nonumber \\
  && \mbox{} 
  + 24 m  L \Lambda^3 \cos\sigma \left( \log {1 \over \Lambda L} 
  \cosh b^\prime- {1 \over 3} b^\prime \sinh b^\prime \right) 
   - {m^2 \over 36 L  \log^2 {1 \over \Lambda L} } (b^\prime)^2 ~.
\end{eqnarray}
It is convenient to introduce dimensionless masses, compactification
scale, and potential:
\begin{equation}
\label{dimlessparameters}
\tilde m \equiv {m \over \Lambda}~, \hspace{0.3cm}
\tilde L \equiv \Lambda L ~, \hspace{0.3cm}
\tilde V \equiv  {L^3 V_{\rm total} \over  48  
                    \tilde{L}^6 \log \tilde{L}^{-1}}\, . 
\end{equation}
The final result for the scalar potential of the mass-deformed SYM theory is:
\begin{eqnarray}
\label{totalpotential2}
\tilde{V} &=& 
  \cosh 2 b^\prime - \cos 2 \sigma 
 + {1\over2}  {\tilde m   \over \tilde{L}^2 } \cos\sigma 
   \left( \cosh b^\prime- \frac{1}{3\log\tilde{L}^{-1}} 
            b^\prime \sinh b^\prime    \right) \nonumber \\
 && \mbox{}
 - {1 \over 1728}  \left({\tilde{m}  \over  \tilde{L}^2}\right)^2 
         {1 \over  \log^3   \tilde{L}^{-1} } \; (b^\prime)^2 ~.
 \end{eqnarray}
 
 The physics that this potential encapsulates is  our main result. 
Before we study the relative importance of the various terms in 
(\ref{totalpotential2}), let us summarize the region of validity of 
the scalar potential. It was derived using weak-coupling semi-classical 
calculations at small $L$, whose validity requires that $\Lambda L \ll 1$. 
The  validity of the fermion-pairing bion mechanism further requires 
$m L \log{1 \over \Lambda L} \ll 1$; the usual soft-breaking condition 
$m \ll \Lambda$ is then automatically satisfied. In other words, both 
dimensionless parameters $\tilde{m}$ and $\tilde{L}$ from (\ref{dimlessparameters}) are small.
 
 We can now use the potential (\ref{totalpotential2}) to study the 
symmetry realization of the theory as the parameters are varied:
\begin{enumerate}
\item Consider the domain:
\begin{equation}
\label{confdomain}   {\tilde{m}\over 8 \tilde{L}^2} \ll 1~,
\end{equation}
In this domain, the bion-induced center-stabilizing term  $(\cosh 2b')$  
dominates over both the monopole and perturbative contributions, both 
of which favor center-symmetry breaking, as we show below. In the regime 
of small ${\tilde{m}\over 8 \tilde{L}^2}$, with  $\tilde{m} > 0$ (recall 
that we set the $\theta$-angle to zero), the vacuum with $\langle \sigma 
\rangle =\pi$, $\langle b^\prime \rangle = 0$ represents the global minimum 
of $\tilde{V}$, while the one with $\langle \sigma \rangle = 0$ is only 
metastable. In this regime, $\langle \tr\, \Omega \rangle= 0$, with unbroken 
center symmetry. To see this explicitly, we expand (\ref{totalpotential2}) 
around   $\langle \sigma \rangle =\pi$, $\langle b^\prime \rangle = 0$, to 
quadratic order in the fluctuations $\delta\sigma$ and $\delta b^\prime$:
\begin{eqnarray}
\label{quadratic}
\tilde{V} &=&
  2 \left(1 - {\tilde{m} \over 8 \tilde{L}^2}
  \left[1 + {2 \over 3   \log \tilde{L}} 
  - {\tilde{m}  \over \tilde{L}^2}{1 \over 432  \log^3 \tilde{L}}
   \right]\right)  \left(\delta b^\prime\right)^2 \nonumber \\
 && \mbox{}
  + 2 \left(1 +  {\tilde{m} \over 8 \tilde{L}^2}\right) 
  \left(\delta\sigma\right)^2 \, ,
\end{eqnarray}
where we have dropped the constant $\tilde V(b^\prime\!=\!0,\sigma\!=\!\pi)$.
In the regime (\ref{confdomain}) there is a mass gap for the dual photon and 
the theory is in the confining phase with unbroken center-symmetry. 

\item
It is clear from (\ref{quadratic}) that, as we depart the regime 
(\ref{confdomain}), center-symmetry becomes destabilized. The leading 
center-breaking effect, in the region where the semi-classical analysis 
is valid, is due to the non-perturbative monopole-instanton term, which 
tends to reduce the mass of the Wilson line. The effect of the 
center-destabilizing perturbative contribution, given by the last 
term in (\ref{totalpotential2}),  is suppressed both numerically  
and parametrically, by the large factor $|\log^3 \tilde{L}|$.

 Thus, keeping $\tilde m$ small and fixed, as we further decrease 
$\tilde L$, the theory leaves the confining domain (\ref{confdomain}). 
The monopole term becomes the most dominant and destabilizes the 
$\langle b^\prime \rangle = 0$ center-symmetric vacuum. As 
eqn.~(\ref{quadratic}) shows, the center-symmetry destabilization 
is continuous. At $\tilde{L}=\tilde{L}_c$ where
\begin{equation}
\label{transition}
 \tilde{L}_c^2  =  {\tilde{m} \over 8} 
  \left[ 1 + {\cal{O}} \left({1\over \log \tilde{L}},{\tilde{m}\over \tilde{L}^2 \log^3 \tilde{L}}\right)
  \right] \, ,
\end{equation}
the center-symmetric vacuum gets destabilized and the two eigenvalues 
of the holonomy smoothly approach each other.
\end{enumerate}
 
 The potential (\ref{totalpotential2}) can be used to study the physics  
until  the semi-classical description breaks down (this occurs when 
the scale of $SU(2) \rightarrow U(1)$ breaking governed by the eigenvalue 
difference $\Delta \theta$ times $1 \over L$ becomes comparable to the 
strong-coupling scale $\Lambda$). The evolution of the eigenvalues leads 
to spontaneous breaking of the $\Z_2$  center symmetry, $\langle \frac{1}{2} 
\tr\,\Omega\rangle=\pm 1$, and the appearance of two vacua.  In the center 
broken phase,  we expect that these two  vacua are continuously connected 
to the two thermal equilibrium states of pure Yang-Mills theory as   
$\tilde m \rightarrow \infty$.

One crucial point here is the following. In the confined phase, the 
effective description of the dynamics is given in terms of the Wilson line 
$\Omega$ (the $b^\prime$-field) and dual photon $\sigma$. On the other 
hand, $\sigma$ is not a well-defined notion in the ``deep" deconfined phase 
where the $SU(2)$ gauge symmetry is fully restored and the abelianization 
of the dynamics is lost. In other words, the  combined potential 
(\ref{totalpotential2}) is strictly valid beyond $L \geq L_c(m)$ and  
for a range  $L \lesssim L_c(m)$ provided the eigenvalues are 
sufficiently apart.  For most of the range  $L < L_c(m)$, the potential 
is solely in terms of $\Omega$, without $\sigma$.

 Finally, we can try to perform a (very rough) extrapolation of our result 
to pure Yang-Mills theory and obtain an estimate of the critical temperature 
of the deconfining phase transition. In the semi-classical domain, from (\ref{transition}), we find 
${T_c \over \Lambda}={1 \over L_c\Lambda} \sim \sqrt{8 \Lambda \over m}$, 
which drops with $m$, but for $m\gg \Lambda $ the result must become 
independent of $m$. Not much is known numerically about the decoupling 
scale for a Weyl fermion in the adjoint representation. In the case of 
$N_c=3$ QCD with three flavors of fundamental fermions it is known that 
relatively large values of the fermion mass, $m\gsim 5\Lambda$, are 
needed in order for the phase transition to approach the deconfinement 
transition of the pure gauge theory \cite{Karsch:2003jg}. Assuming that 
the decoupling scale for an adjoint Weyl fermion is in the range
$m_{\it dec} \sim (5-10)\Lambda$ we expect $T_c\sim (0.8-1.3)\Lambda$, 
broadly consistent with lattice data.

\section{Pure Yang-Mills theory}

 In the previous Section, we showed that for  $\tilde{m}\ll 1$ the 
center-symmetry restoring phase transition can be described semi-classically.
In this regime the transition is driven by the competition between 
center-stabilizing topological molecules and center-destabilizing 
monopole-instantons. In this Section, we will show that the same
mechanism also exists in the pure gauge theory, even though in this 
case the effects cannot be computed reliably. This implies that it
is plausible that the deconfinement transition in pure gauge theory
is driven by the same topological phenomena that operate in the 
small $\tilde{m}$ limit. 

\subsection{Non-perturbative effects on the classical background}

 In this Section, we will consider possible non-perturbative contributions 
to the potential for the Wilson line in pure Yang Mills theory. The question 
is whether there are terms that favor the center-symmetric vacuum and
compete with the perturbative contributions to $V(\Omega)$. We consider
a {\it classical} background field on $\R^3 \times \S^1$: 
\begin{eqnarray}
\Omega= \left( \begin{array}{cc} 
e^{i \Delta \theta/2} &     \\
  &  e^{- i \Delta \theta/2} 
\end{array} \right)\, ,
\label{bcgr}
\end{eqnarray}
where $\Delta \theta$ is the separation between the eigenvalues of Wilson 
line. In the classical background  (\ref{bcgr}), and at weak coupling, the  
Wilson line behaves as an adjoint Higgs field breaking the microscopic 
$SU(2)$ symmetry down to $U(1)$ at large distances. As before there are 
two types of elementary monopole-instantons, ${\cal M}_{\rm 1}$ and  
${\cal M}_{2}$. The amplitudes associated with these instanton events 
are essentially the ones given by (\ref{1amplitudes}), but now without 
the fermion zero modes:
\begin{eqnarray}
\begin{array}{ll}
 {\cal M}_{\rm 1} = e^{- \frac{4 \pi}{g^2} \Delta \theta +  i \sigma} 
    \equiv  e^{-b + i\sigma},  \quad\quad &
 \overline {\cal M}_{ 1} = 
     e^{ - \frac{4 \pi}{g^2} \Delta \theta  - i \sigma}  \equiv  e^{-b -  i\sigma}, 
 \\[0.1cm]  
 {\cal M}_{2} =   e^{- \frac{4 \pi}{g^2} (2 \pi -\Delta \theta) -   i \sigma}  
    \equiv   \eta e^{+b - i\sigma},   \qquad &
 \overline {\cal M}_{2 } =  e^{ - \frac{4 \pi}{g^2} (2 \pi- \Delta \theta) +   i \sigma}  
  \equiv   \eta  e^{+b + i\sigma}~.
\end{array}   
\label{nonsusyMs}
\end{eqnarray}
The interaction between different monopole-instantons with magnetic 
charge $n_m$ and scalar charge $n_b$ can be computed by using the 
two point correlator, as in (\ref{qzm1}):
\begin{eqnarray}
\langle  e^{n_b b +  i n_m \sigma} ({\bf x})  
          e^{ \bar n_b b +  i \bar n_m \sigma} ({\bf y}) \rangle_0 
=  e^{-V(|{\bf x-y}|) } 
=  e^{-  \frac{4 \pi L}{g^2 |{\bf x-y}| } (-n_b \bar n_b+ n_m \bar n_m) }~.
 \end{eqnarray} 
This formula is true classically. At weak coupling, the $b$ field may 
acquire a radiatively induced mass. This modifies the potential as:
\begin{equation} 
V(|{\bf x-y}|)  =   \frac{4 \pi L}{g^2 |{\bf x-y}| } 
 (-n_b \bar n_b e^{-m_b|{\bf x-y}|} + n_m \bar n_m) ~.
\label{modified}
\end{equation}
In perturbation theory, there are three possibilities for $m_b^2$ (recall 
that $m_b$ is the mass for the Wilson line,  $m_b^2 |\tr \Omega|^2$,  
obtained by studying small fluctuations around the classical vacuum): 
{\it i)} In supersymmetric theories with supersymmetry preserving boundary 
conditions,  $m_b^2=0$ to all orders in perturbation theory. 
{\it ii)} If $m_b^2>0$, as is the case for QCD(adj) with periodic boundary 
conditions for fermions \cite{Unsal:2007vu,Unsal:2007jx}, then the potential 
is as in (\ref{modified}). In particular, the $b$-exchange interaction is 
short range and has no effect on the long distance effective theory. 
{\it iii)} If $m_b^2<0$, as in thermal YM, then there exists an instability 
of the center-symmetric vacuum.  

 For our purposes we will consider $m_b=0$, because classically there is 
no mass gap for the holonomy fluctuations. We can now write down the 
effective theory for an ensemble of monopole-instantons. We get:
\begin{equation}
L=  {1 \over 2} {g^2\over (4 \pi)^2 L}
 \left[(\partial_i b)^2 + (\partial_i \sigma)^2\right] - 
   (e^{-b} + \eta e^{+b})  \cos \sigma  + \ldots,
\label{leff1} 
\end{equation}
where we have not attempted to determine the overall coefficient of 
the effective potential, and we defined $b \equiv \frac{4 \pi}{g^2} 
\Delta \theta$. The ellipsis denote both perturbative and other 
non-perturbative contributions. The effective potential in (\ref{leff1}) 
arises by summing  the contributions of the different monopole-instantons in (\ref{nonsusyMs}):
\begin{eqnarray}
- V^{\rm n.p. }(\Delta \theta, \sigma) &=& 
 (  {\cal M}_{\rm 1} + {\cal M}_{2} +  \overline {\cal M}_{2 }  
      + \overline {\cal M}_{2 } ) \cr  \cr 
& = &  2  \left[  e^{- \frac{4 \pi}{g^2} \Delta \theta  }  
      +   e^{- \frac{4 \pi}{g^2} (2 \pi -\Delta \theta) } \right]   \cos \sigma 
  =  4   e^{- \frac{4 \pi^2}{g^2}} \cosh b^\prime  
      \cos \sigma~.
\label{np1}
\end{eqnarray} 
We would like to make a number of comments regarding this potential: 

{\bf 1)} We note that the potential is not quite sensible as it is not 
manifestly periodic in $\Delta \theta$. This problem can be addressed
by taking into account the existence of an infinite tower of 
monopole-instantons, see below. 

{\bf 2)} We observe that the potential has an extremum at $\sigma=0$ and 
$\Delta \theta= 0$ where the semi-classical approximation is not reliable. 
For the stability of the center-symmetric vacuum we must have a global 
minimum at  $\Delta \theta= \pi$, and the sigma field must be non-tachyonic 
there. However, around  $(\Delta\theta,\sigma)= (\pi,0)$, and also around  
$(\Delta\theta,\sigma)=(\pi,\pi)$, the expansion of the potential in 
terms of small fluctuation yields:
\begin{eqnarray}
V^{\rm n.p. }(b', \sigma) 
=  2  e^{- \frac{4 \pi^2}{g^2}} 
\left\{ \begin{array} {ll}
   - (\delta b')^2  + (\delta\sigma)^2 + \ldots\;\;  &  {\rm for}\; 
  (\Delta\theta,\sigma)= (\pi,0) \\[0.1cm]
   + (\delta b')^2  - (\delta\sigma)^2 + \ldots\;\;  &  {\rm for}\;
  (\Delta\theta,\sigma)= (\pi,\pi)
\end{array} \right. \, . 
\label{fluc1} 
\end{eqnarray} 
At leading order in the semi-classical expansion, the Hessian around 
each  of the two  center-symmetric saddle points is negative and 
we conclude that monopole-instanton effects do not favor the 
center-symmetric vacuum.

  In the following we will address both of these points. We will 
argue, in particular, that topological molecules can stabilize 
the center. 

{\bf 1)} {\it Making the potential periodic:}
 Because the $\Delta \theta$ field is an angular variable there is an
infinite set of monopole-instantons on $\R^3 \times \S^1$. For magnetic 
charge $+1$, there is a tower of monopole-instantons with topological 
charges $Q_T (n_w) ={\Delta\theta \over 2 \pi} + n_w, n_w\in \Z$. We observe 
that ${\cal M}_{\rm 1}$ and $\overline {\cal M}_{2 } $ are the $n_w=0$  
and $n_w=-1$ members of this tower. Following \cite{Poppitz:2011wy}
we define the generalized fugacity of the monopole-instanton amplitude 
as:
\begin{eqnarray}
F(\Delta \theta) e^{i \sigma} 
   =  \left[ \sum_{n_w \in \mathbb Z} 
        e^{- \frac{4 \pi}{g^2} |\Delta \theta + 2 \pi n_w|} \right]  
        e^{  i \sigma},
\end{eqnarray}
Clearly, $F(\Delta \theta + 2 \pi) =F(\Delta \theta)$ is a periodic 
function. Using Poisson resummation it can be rewritten as: 
\begin{eqnarray}
\nonumber
 F(\Delta \theta ) = \frac{1}{\pi}  \sum_{n_e \in \mathbb Z}  
 \frac{\frac{4 \pi}{g^2}} { \left(\frac{4 \pi}{g^2} \right)^2 + n_e^2} 
    e^{i n_e \Delta \theta } 
\equiv \frac{1}{\pi} \left( 1+ 2  \sum_{n_e =1}^{\infty}  c_{n_e}
   \cos n_e \Delta \theta  \right), ~ 
c_{n_e} \equiv 
    {{4 \pi \over g^2} \over \left(4 \pi \over g^2 \right)^2 + n_e^2}~,
\end{eqnarray}
which, after using $\cos n_e \Delta \theta = \half |{\rm tr} \Omega^{n_e}|^2  
-1$, can be expressed  in terms of the gauge invariant Wilson line:
\begin{eqnarray}
\label{dyon} 
F(\Omega) =  \frac{1}{\pi} \left( 1+   \sum_{n_e =1}^{\infty}  c_{n_e}
   (|{\rm tr} \Omega^{n_e}|^2  -2) \right)~.
\end{eqnarray}
The non-perturbatively induced potential, incorporating the entire 
Kaluza-Klein tower of the monopole-instantons takes the form: 
\begin{eqnarray}
V^{\rm n.p. }(\Omega, \sigma) && =  F(\Omega)  \cos \sigma~.
\label{np2}
\end{eqnarray} 
The extremum of (\ref{np2}) is at $\Delta \theta=0, \sigma=\pi$. At 
this point center-symmetry is broken, gauge symmetry is restored and 
the abelian long distance description is invalid, exactly as for 
(\ref{np1}). Thus, the leading-order bosonic potential induced by 
monopole-instantons does not appear to stabilize center symmetry.

{\bf 2)} {\it Role of topological molecules}:
We showed that in $\N=1$ SYM\footnote{The same is true in $\N=2$  
Seiberg-Witten theory softly broken to $\N=1$.} monopole-instantons
carry fermionic zero modes and do not induce a bosonic potential. 
Instead, monopole-instantons generate a fermion mass term. A bosonic 
potential appears at second order in the semi-classical expansion through 
the terms induced by composites $[{\cal M}_{\rm i}\overline{\cal M}_{\rm j}]$.  
The magnetic bion combinations $[{\cal M}_{\rm 1}\overline{\cal M}_{2}] 
+ [{\cal M}_{2}\overline{\cal M}_{\rm 1}]$ induce a mass gap for gauge 
fluctuations, and the center stabilizing bions $ [{\cal M}_{\rm 1} 
\overline{\cal M}_{\rm 1} ]+ [{\cal M}_{2}\overline{\cal M}_{2}]$ 
stabilizes the center-symmetric vacuum. We may therefore ask whether 
topological molecules induce similar effects in pure Yang-Mills theory.

  Symmetry permits, at second order in the  semi-classical expansion, 
terms of the form:
\begin{eqnarray}
-V^{(2)}(b, \sigma) =   c_1  \eta  \cos 2\sigma 
 + c_2  \left(  e^{- 2b } + \eta^2  e^{ 2b } \right)  
 + c_3  \left(  e^{- 2b} + \eta^2  e^{ 2b } \right)  \cos 2 \sigma 
 + c_4  \eta 
\label{np3}
\end{eqnarray}
The terms in (\ref{np3}), which can be made periodic in $b$ by a procedure 
similar to the one that led to (\ref{np2}), can be thought as due to  
molecular monopole-instantons of the form:
\begin{eqnarray}
\begin{array}{ll}
  [{\cal M}_{1}\overline {\cal M}_{ 2} ]  \sim  e^{+2i\sigma}~, \qquad 
& [{\cal M}_{2}\overline {\cal M}_{ 1} ]  \sim  e^{-2i \sigma}~, \cr  
  [{\cal M}_{1}\overline {\cal M}_{ 1} ]  \sim  e^{- 2b}~, \qquad 
& [{\cal M}_{2}\overline {\cal M}_{ 2} ]  \sim  \eta^2  e^{+ 2b}~, \cr 
  [{\cal M}_{1}{\cal M}_{ 2} ] \sim \eta~, \qquad   
& [\overline {\cal M}_{1} \overline {\cal M}_{ 2} ] \sim \eta~,   \cr 
  [{\cal M}_{1}{\cal M}_{ 1} ] \sim e^{- 2b+ 2 i \sigma }~,   \qquad  
& [\overline{\cal M}_{1}\overline {\cal M}_{ 1}] \sim e^{- 2b- 2i\sigma}~, \cr 
  [{\cal M}_{2}{\cal M}_{ 2} ] \sim  \eta^2 e^{+ 2b -  2 i \sigma }~,   \qquad 
& [\overline{\cal M}_{2}\overline {\cal M}_{ 2}] \sim   
                                           \eta^2 e^{+ 2b +  2i\sigma}~.
\end{array}
\label{list}
\end{eqnarray}
In Section \ref{sec:centersymmetry}, we showed that there are SUSY and 
softly broken SUSY theories in which we can reliably demonstrate that 
these molecules generate a center-symmetric minimum. Below we will 
demonstrate, using the BZJ prescription, that topological molecules
also stabilize the center in pure YM theory. Clearly, in order for 
the second order terms in (\ref{np3}) to be more important than the 
leading-order term in (\ref{np2}), we have to push the expansion
beyond the semi-classical domain. We will therefore not attempt 
to perform a quantitative calculation. Our goal is to show that 
the same mechanism that drives the center-symmetry changing 
transition in the softly broken $\N=1$ theory also operates 
in pure Yang-Mills theory. 

\subsubsection{Quasi-zero modes and  bion amplitudes }

 In this Section, we will study the amplitude of topological molecules
in the pure Yang-Mills theory on the classical background. The amplitudes 
for the molecular monopole-instanton events can be found by integrating 
over the relative separation quasi-zero mode between its constituents. 
This will be similar to the study in Section \ref{sec:centersymmetry}, 
where we performed an analysis for the case of $n_f$ adjoint fermions, 
where $n_f=1$ corresponds to $\N=1$ SYM theory. The result for pure 
Yang Mills theory can be obtained by taking the limit $n_f= \epsilon 
\rightarrow 0$ in  (\ref{m-bion}), (\ref{naive}), and (\ref{sign}). 

 We begin with topological molecules of the type $[{\cal M}_{1}  
\overline {\cal M}_{ 2}]$.  The integral over the quasi-zero mode is 
the same as in (\ref{m-bion}), except that we have to take the $\epsilon 
\rightarrow 0$ limit. We find
\begin{eqnarray} 
I(\lambda, \epsilon) 
 =  \lambda^{4 \epsilon -3} \Gamma(4\epsilon-3) 
 = \lambda^{-3}  \left( - \frac{1}{24\epsilon}  
       + \frac{1}{6} \left(- \log(\lambda) + \gamma  -\frac{11}{6} \right) 
       + O(\epsilon) \right)\, . 
\label{m-bion2}
\end{eqnarray}
The divergence in the $\epsilon \rightarrow 0$ limit is due to 
over-counting of uncorrelated monopole-instanton events, which are already 
included in the dilute monopole-instanton gas approximation. In theories 
with fermions, this long-distance divergence is cut-off by the fermion 
zero mode exchange, both in quantum mechanical examples \cite{Balitsky:1985in}
and in quantum  field theories \cite{Poppitz:2011wy, Argyres:2012}.
In theories without fermions, in order not to double-count, this divergence 
needs to be subtracted, see \cite{Schafer:1996wv, Bogomolny:1980ur} for 
a quantum mechanical example. Consequently, the  prefactor of the 
magnetic bion amplitude is:
\begin{eqnarray}
c_1(g)  = \frac{2 \pi  a^2}{3}   \left(\frac{8\pi}{g^2}\right)^3  
 \left(- \log\left (\frac{g^2}{8 \pi}\right) +\gamma-\frac{11}{6} \right)\, ,
\label{c1}
\end{eqnarray} 
where $a$ is the coefficient of the one monopole-instanton amplitude, 
which is set to one in (\ref{nonsusyMs}) and which can be restored if 
desired. 

 Next, we consider the neutral bions $[{\cal M}_{1}\overline{\cal M}_{1}]$. 
Here, the 
constituents interact attractively both due to $\sigma$ and $b$-exchange, 
and we need to apply the BZJ prescription. The result is:
\begin{eqnarray}
{c}_2(g) = \frac{2 \pi  a^2}{3}  \left(-\frac{8\pi}{g^2}\right)^3     
 \left(- \log\left (-\frac{g^2}{8 \pi}\right) +\gamma-\frac{11}{6} \right)  
= - c_1(g) \pm  (i \pi) \frac{2\pi a^2}{3} \left(\frac{8\pi}{g^2}\right)^3 \, . 
\label{c2}
\end{eqnarray} 
In analogy with the quantum mechanics example where the ambiguity 
associated with non-Borel summability of the perturbation theory is 
canceled by the  molecular instanton-anti-instanton  contribution 
\cite{Bogomolny:1980ur,ZinnJustin:1981dx}, we also expect the 
ambiguity associated with the non-Borel summability of the pure YM 
theory on $\R^3 \times S^1$ to be canceled by the two-fold ambiguity 
of the amplitude for neutral bions. This implies that the imaginary 
part in (\ref{c1}) must cancel by  large-orders in perturbation theory, 
and we discard that term. The remaining term in (\ref{c2}) satisfies
\begin{eqnarray}
 {c}_2(g) = - c_1(g)~.
\label{c2-add}
\end{eqnarray} 
For a more detailed discussion of this conjecture and the available  
theoretical evidence in its favor, see \cite{Argyres:2012}. 

Next, consider the  $[{\cal M}_{1}{\cal M}_{ 1}]$ and $[{\cal M}_{2}
{\cal M}_{ 2}]$ amplitudes. Now, $\sigma$ exchange is repulsive and $b$
exchange is attractive, leading to a cancellation. This means that the self-dual 
monopole-instanton events are not correlated, and the molecular amplitude
vanishes, ${c}_3(g)=0$. The same conclusion is also obtained in 
Ref.~\cite{Diakonov:2007nv}. 

 For the $[{\cal M}_{1}{\cal M}_{ 2}]$ amplitude, slightly more care is needed. The 
interaction again cancels between attractive $\sigma$ exchange and 
repulsive $b$ exchange. This would seem to imply $c_4(g)=0$, as in 
the previous case. However, $[{\cal M}_{1}{\cal M}_{ 2}]$ molecules
with sizes less than the compactification scale correspond to ordinary  
BPST-instantons. Taking into account small 4d instantons corresponds to 
$c_4(g) \sim \eta$. Large 4d instantons do not exist because large 
uncorrelated $[{\cal M}_{1}{\cal M}_{ 2}]$ molecules are already 
included in the instanton-monopole contribution. This implies that there 
is a finite instanton term, but since the 4d-instanton does not depend 
on the $b$ and $\sigma$ field we find that the leading semi-classical 
instanton contribution only enters as a constant term in the effective 
action. 

 Combining the terms (\ref{c1}), (\ref{c2}), and (\ref{c2-add}) that appear 
at second order in the semi-classical 
expansion we find:
\begin{eqnarray}
V^{(2)}(b, \sigma) =   - c_1 \eta  \cos 2\sigma + 
 c_1   \left(  e^{- 2b } + \eta^2  e^{ 2b } \right)~.  \label{np4} 
\end{eqnarray}
This is indeed the same result as in (softly broken) $\N=1$  supersymmetric 
theory. The combined potential of monopole-instantons and bions is 
\begin{eqnarray}
V^{(1)}(b, \sigma) + V^{(2)}(b, \sigma) 
=  - \eta^{1/2} \cosh b' \cos \sigma   
   - c_1 \eta  \cos 2\sigma 
   + c_1  \eta  \cosh 2 b'   \, , 
\label{np5} 
\end{eqnarray}
leading to: 
\begin{eqnarray}
 \tilde m_{b'}^2  &=&  \left[ - \eta^{1/2} +  4 c_1 \eta  \right]~,  \\[0.1cm] 
 \tilde m_{\sigma}^2  &=& \left[  \eta^{1/2} +  4 c_1 \eta  \right]~.
\label{masses3}
\end{eqnarray}
Thus, we find that second-order effects in the semi-classical expansion, in 
particular neutral bions, stabilize the center symmetry, without destabilizing 
the $\sigma$ mode. As in the discussion around (\ref{dyon}) and (\ref{np2}),  
in order to make periodicity of the potential manifest, we may  sum over 
the Kaluza-Klein tower of the neutral bions. This leads to a center-stabilizing potential 
for the Wilson line, given by  $V_{\rm b-tower}  (\Omega) \sim  \sum_{n}    
|{\rm tr} \Omega^{n}|^2 $, similar to  (\ref{dyon}). However, as stated 
earlier, the stabilization of center-symmetry by the neutral bion induced potential 
requires strong coupling  where neutral bion term can  overcome the 
monopole-instanton induced instability as well as the perturbatively 
induced  potential. In this regime,  we cannot perform a quantitative 
calculation. It is nevertheless
intriguing that the same effect that leads to center stabilization at weak
coupling for softly broken $\N=1$ theory is  present in pure Yang Mills
theory as well. For an attempt to connect semi-classical neutral bion 
molecules to strong coupling effects and infrared renormalons,  see the recent work \cite{Argyres:2012}.

\subsubsection{Connecting monopole-instantons to dyon particles}

 The neutral bion induced potential (\ref{np4}), just like the 
monopole-instanton induced potential, is not manifestly periodic in 
$\Delta \theta$. This problem, as in the case of monopole-instantons,  
can be addressed by taking into account the whole Kaluza-Klein tower 
of neutral bion molecules. In this Section, we will show that the 
tower of monopole-instantons can be interpreted, via Poisson resummation,
as the contribution from dyon particles.\footnote{A number of authors, see, e.g., \cite{Diakonov:2010tz} and references therein, refer to the monopole-instantons ${\cal M}_{1}$ and ${\cal M}_{2}$ as ``dyons", because 
they are magnetically charged and self-dual. We believe that this choice 
of words is not quite appropriate. The amplitude of a  monopole-instanton has 
the form $e^{-b + i\sigma} \sim e^{-A_4 + i \sigma}$. However, a dyon 
particle with both electric and magnetic charge couples to $(A_4,\sigma)$ 
as $e^{i q_e A_4+ i q_m  \sigma}$, in both Minkowski and Euclidean space. 
We will see that the Kaluza-Klein tower of monopole-instantons can 
be written as a sum over dyons that exhibit the expected coupling
for electrically and magnetically charged particles.}

Consider the Poisson duality relation for the tower of monopole-instantons, 
see (\ref{dyon}):
\begin{eqnarray}
\left[ \sum_{n_w \in \mathbb Z} 
  e^{-\frac{4 \pi}{g^2} |\Delta \theta + 2\pi n_w|} \right]  e^{i\sigma}  
= \left[  \frac{1}{\pi} \sum_{n_e \in \mathbb Z}  
 \frac{\frac{4 \pi}{g^2}} { \left(\frac{4 \pi}{g^2} \right)^2 + n_e^2} 
  e^{i n_e \Delta \theta } \right]  e^{i \sigma}  \, . 
  \label{duality}
 \end{eqnarray}
The phase  $e^{i n_e L A_4 }= e^{i n_e \Delta \theta }$ is the canonical 
coupling of a charged particle to a background gauge field $A_4$,  and  
$e^{  i \sigma}$ is a 3d instanton amplitude associated with magnetic 
charge one. 

Consider a massive particle on $\R^{3,1}$ with  magnetic and electric 
charge $(n_m, n_e)$ and mass $M_{(n_m, n_e)}$, called a dyon.  By 
Gauss' law,  there is a combined electric and magnetic  flux coming 
out of this particle;
 \begin{eqnarray}
 \int_{\S^2 = \partial \R^3} (\vec E + i \vec B)\cdot d \vec \Sigma 
=  n_e + i \frac{4 \pi}{g^2} n_m 
\equiv q_e + i q_m\, , 
  \end{eqnarray}
where $\S^2= \partial \R^3$ is a sphere at infinity. We can analytically 
continue to Euclidean time  (consider the theory on $\R^{4}$) and then  
compactify one of the directions, i.e.,~consider the theory on  
$\R^{3} \times \S^1$. 

 A dyon particle whose world-line wraps the $\S^1$ corresponds to a 
finite action instanton, $S_{(n_m, n_e)} = LM_{(n_m, n_e)}$.  The 
amplitude associated with the dyon is, 
\begin{equation}
e^{-S_{(n_m, n_e)} } e^{i n_e \theta_e  + i n_m \theta_m} 
 \equiv  e^{-LM_{(n_m, n_e)} } e^{i n_e \theta_e  + i n_m \sigma} \, , 
\end{equation}
where  $e^{i \theta_e}$ and $e^{i\theta_m}$ are electric and magnetic 
Wilson lines. The magnetic Wilson line is naturally interpreted in 
terms of the dual photon\footnote{We can see this starting from
abelian duality in 4d, then compactify the theory on $\R^3 \times 
\S^1$, and finally match the result to 3d abelian duality. The abelian 
duality on 4d is expressed by $F_{\mu \nu}^D =\half \epsilon_{\mu\nu 
\rho\sigma} F_{\mu \nu}$ where $F_{\mu \nu} = \partial_{\mu} A_{\nu} 
- \partial_{\nu} A_{\mu}$ and $F_{\mu \nu}^D = \partial_{\mu} A_{\nu}^D 
- \partial_{\nu} A_{\mu}^D$, where $A_\mu,A_\mu^D$ are the original 
and the dual gauge potential. Using dimensional reduction and splitting 
the duality relation into $4i$ and $ij$ components,  we obtain 
$\partial_i A_4^D = -\half \epsilon_{ijk} F_{jk}$ and $F_{ij}^D=  
\epsilon_{ijk} \partial_k A_4$. The first relation is the well-known 
abelian duality relation in 3d, where we identify  $A_4^D = \sigma 
\equiv \theta_m$ and $A_4=  \sigma^D = \theta_e$.  The monopole-instanton 
amplitude is naturally expressed in terms of $\theta_m= \sigma$ and
$\theta_e \equiv \Delta \theta $.} $\sigma$.
 The  duality relation (\ref{duality}) therefore admits an interesting  
interpretation: its right hand side may be re-written as a sum over 
massless/light dyons with charges $(n_m, n_e) = (1, n_e)$ whose world-lines   
wrap around the $\S^1$:
\begin{eqnarray}
 \frac{1}{\pi} \sum_{n_e \in \mathbb Z}  
  \frac{q_m}{ q_m^2 + q_e^2} e^{-LM(q_m, q_e)}  
  e^{i n_e  \theta_e  + i  \sigma}~.
\end{eqnarray}
In the classical background, the long distance dynamics is abelian 
and the sum over monopole-instanton amplitudes maps, via Poisson 
resummation, to a sum over the electric charges of the dyons. 

 From this point of view, we find (\ref{duality}) quite intriguing. 
This relation makes perfect sense in  $\N=2$ Seiberg-Witten theory \cite{Poppitz:2011wy}. 
In that context, on $\R^4$, the theory has a classical moduli space along 
which gauge symmetry is reduced to $U(1)$ due to adjoint Higgsing by 
the vev $\langle \Phi \rangle = v T^3$. In the semi-classical domain of the  
$\N=2$ theory, the Poisson duality relation is reliable. Moreover, in  
$\N=2$ theory, the combination  $\int_{\S^2 } v (\vec E + i \vec B)\cdot
d \vec\Sigma = Z_{(n_m, n_e)}$ appears as the central charge in the 
supersymmetry algebra.

 In pure Yang-Mills theory, the combination  of electric and magnetic 
charges $\int_{S^2 } (\vec E + i \vec B)\cdot d \vec\Sigma $ (without 
the vev) appears naturally. However, we also know that there is no apparent 
adjoint Higgsing on $\R^4$ and it seems hard to make sense out of the 
Poisson duality relation in that case. One can speculate that there may 
be a connection between Poisson duality and the maximal abelian gauge 
proposal by 't Hooft \cite{'tHooft:1981ht}, which is based on using a 
specific gauge condition applied to a composite operator to define an 
adjoint Higgs field, and the corresponding abelianized $SU(N)\rightarrow 
U(1)^{N-1}$ theory. In that case, of course, monopoles (as well as dyons) 
appear. However, the theory is strongly coupled, and the abelian dynamics 
is not well understood. This is a direction worthy of further pursuit.

\section{Comments on the literature}

 In this Section, we will attempt to clarify the relation between our 
work and previous analytical approaches to the deconfinement transition in the literature. 

The question whether 
one can study the deconfinement phase transition at weak coupling 
was considered by  Aharony et al.~in \cite{Aharony:2005bq}. These
authors found a realization of this idea in a finite spatial volume
$\S^3 \times \S^1$, but rejected the possibility that an example
can be found for an infinite spatial volume, in particular on 
$\R^3 \times \S^1$.  Ref.\cite{Aharony:2005bq} studied large-$N$ 
(strictly, $N=\infty$) pure Yang-Mills theory on small $\S^3 \times 
\S^1$ by integrating out the perturbatively weakly coupled modes 
and thus mapping the field theory to a matrix model. This approach 
pushes the deconfinement transition to the weak coupling regime. 
However, since they study gauge theory on a small sphere $\S^3$, 
approaching the thermodynamic limit requires taking the infinite-$N$ 
limit.  In particular, the approach of \cite{Aharony:2005bq} does not 
apply to gauge theories of finite rank. 

 A way around the obstacle of analytical ``intractability"  of 
deconfinement in an infinite spatial volume was found in 
Refs.~\cite{Simic:2010sv,Anber:2011gn} by compactifying the gauge 
theory on $\R^2 \times \S^1_L \times \S^1_\beta$ where $\S^1_L$ is 
a spatial circle and  $\S^1_\beta$ is the thermal circle. The size 
of the spatial circle provides a tunable control parameter, the 
counter-part of $\S^3$ in the approach of \cite{Aharony:2005bq}. 
The advantage of this formalism is that the small volume theory 
is still a field theory  in an infinite spatial volume (as opposed 
to a matrix model), although it is  $\R^2$ and not $\R^3$.  The  
approach of \cite{Simic:2010sv, Anber:2011gn} also works for 
finite rank gauge theories, mapping the deconfinement transition  
to a  phase transition in two-dimensional spin systems; the relevant 
spin systems are  the affine XY-spin models with symmetry breaking 
perturbations. 

 In the present paper, we gave a reliable semi-classical analysis 
of the center-symmetry changing phase transition on $\R^3 \times 
\S^1$. The use of semi-classical methods in connection with the 
center-symmetry changing transition was previously investigated 
in a series of papers by Diakonov and collaborators
\cite{Diakonov:2002fq,Diakonov:2007nv,Diakonov:2009jq,Diakonov:2010tz}:

{\bf 1)} The first of these papers, a review published in 2002
\cite{Diakonov:2002fq}, suggests that center-symmetry can be stabilized 
by monopole-instantons (dyons, in the language of Diakonov et al.). 
Diakonov obtains the term $V \sim \cosh \frac{4 \pi}{g^2} (\pi 
-\Delta \theta)$ in the monopole-instanton induced potential, 
see (\ref{monopolepotential}) and (\ref{np1}), and observes
that this contribution has a minimum at the center symmetric point $\Delta 
\theta= \pi$. However, this conclusion is based on neglecting the 
the $\cos \sigma$ term which arises from the coupling to the dual 
photon. Indeed, we found that in the semi-classical case $m\ll\Lambda$
monopole-instantons favor center-symmetry breaking. 

{\bf 2)} $\N=1$ SYM theory on $\R^3 \times \S^1$ is presented 
as an example for center stabilization by  monopole-instantons
in Refs.~\cite{Diakonov:2002fq,Diakonov:2007nv, Diakonov:2010tz}.
This interpretation is common in the literature but, as we 
showed above, it is physically not correct. Monopole-instantons 
in  $\N=1$ SYM have fermion zero modes, and they do {\it not} generate 
a bosonic potential for the Wilson line. Rather, they generate a 
fermion bilinear which determines the superpotential. The bosonic 
potential can be found via supersymmetry using  $V \sim 
|\frac{\partial W}{\partial b}|^2$, as in Section \ref{susy}, or 
using the BZJ-prescription, as in Section \ref{susywithoutsusy}.
In either case the conclusion is that the physical mechanism
 generating the potential for the Wilson line is related 
to topological molecules. 

{\bf 3)} A quantitative theory of a confining ensemble of 
monopole-instantons was proposed in \cite{Diakonov:2007nv}.
The paper argues that one can analyze the vacuum of Yang-Mills 
theory by semi-classical means, and that, furthermore, the only 
topological configurations that contribute to the non-perturbative 
potential for the gauge holonomy satisfy: {\it i)} the self-duality  
condition, and   {\it ii)} a (magnetic) charge neutrality constraint.   
This means that, contrary to item {\bf 1)}, instanton-monopoles 
$[{\cal M}_1]$, $[{\cal M}_2]$ and $[{\overline{\cal M}}_1]$, 
$[\overline{\cal M}_2]$ are excluded. The self-dual neutral objects 
are KvBLL-calorons   \cite{Lee:1997vp,Kraan:1998sn}, 4d BPST instantons 
at finite temperature and non-zero holonomy, and multi-calorons. 
Calorons can be viewed as topological molecules of the form $[{\cal M}_1
{\cal M}_2]$ and $[\overline{\cal M}_1\overline{\cal M}_2]$. We observed
in (\ref{list}) that at the classical level there is no coupling of 
instantons to the holonomy. The claim in Ref.~\cite{Diakonov:2007nv} 
is that a potential for the Polyakov line can arise from the collective 
coordinate measure. This contribution is formally of higher order in 
the coupling constant.

  In the pure gauge theory the critical temperature for 
deconfinement is of order $\Lambda$, and it is clear that there 
cannot be a systematic semi-classical theory of the transition. The 
continuity argument outlined in Figure \ref{fig:phase2} provides a 
less ambitious program: We connect the strongly coupled center-symmetry 
changing phase transition in Yang-Mills theory to a semi-classical
phase transition in mass-perturbed super-Yang-Mills theory on 
$\R^3 \times \S^1$. The semi-classical calculation at small $m$ provides 
two important lessons: {\it a.}) Both monopole-instantons (self-dual and 
magnetically charged objects\footnote{It is clear that there cannot 
be a general argument that rules out contributions from magnetically
charged objects. In particular, Polyakov's solution to the Yang-Mills 
adjoint Higgs system on $\R^3$ \cite{Polyakov:1976fu} and the solution 
of deformed Yang-Mills theory \cite{Unsal:2008ch} on  $\R^3 \times \S^1$ 
map the gauge theory partition function to a grand canonical ensemble of  
magnetic charges.}) and neutral bions (non-self dual and magnetically 
neutral objects like $[{\cal M}_1\overline{\cal M}_1]$) contribute to 
the potential for the gauge holonomy and {\it b.}) these are the leading 
contributions to the potential in the controllable small-$L, m$ regime. 
Thus, we believe that our results contradict the assumptions 
in~\cite{Diakonov:2007nv}.

 Finally, we also note that detailed phenomenological studies of 
the effective lagrangian for the Polyakov line in pure Yang-Mills 
theory were carried out by Pisarski and collaborators \cite{Dumitru:2003hp,Pisarski:2006hz,Dumitru:2010mj,Dumitru:2012fw}.  
Center-stabilizing double-trace operators are considered  in 
Ref.~\cite{Unsal:2008ch}  to address a semi-classical mechanism of confinement and large-$N$ volume independence, and in  Ref.~\cite{Myers:2007vc} to study phases with partial center symmetry breaking.   The present study provides a microscopic explanation of the origin of 
the center-stabilizing  double-trace terms in these studies. 

\section{Conclusions and outlook}

 In this paper, we argued that the center-symmetry changing phase 
transition in thermal Yang Mills theory is continuously connected 
to the phase transition in softly broken $\N=1$ theory on 
$\R^3 \times \S^1$. We showed that for small values of the 
adjoint fermion mass $m$ the critical scale $L_c$ is analytically 
calculable. We also provided theoretical evidence that the 
same mechanism that drives the phase transition at small $m$, 
the competition between center-stabilizing topological molecules
and center-destabilizing monopole-instantons,  also exists
in pure Yang Mills theory.  

 There are a number of directions worthy of further pursuit, 
including both  numerical and analytical studies: 
\begin{enumerate}
\item 
The phase diagram in Figure \ref{fig:phase2} can be explored with 
lattice methods that are available today. Furthermore, lattice simulations  
can be used to study the role of neutral non-self-dual topological defects 
(such as the neutral bions), for example, by using the techniques of  
Ref. \cite{Bruckmann:2011cc}. 

\item
One may generalize the semi-classical study of center-symmetry changing 
phase transition on $\R^3 \times \S^1$ to all gauge groups, exploring the 
role of various topological excitations in the symmetry realization and 
the nature of the phase transition. 
\end{enumerate}
Let us also make a few comments on the $SU(N)$, $N\geq 3$ case.  Here, 
there are $N$-types of monopole-instantons, associated with the $(N-1)$ 
simple roots and one affine root. As in the $SU(2)$  theory,  in the softly 
broken $\N=1$ theory, these monopole-instantons, as well as perturbative 
one-loop fluctuations, tend to break center-symmetry. There are $N$-types 
of magnetic bions and   $N$-types of neutral  bions. The magnetic bions, 
for all $N\geq 3$, unlike the $N=2$ case,  lead to a center-destabilizing 
potential. As in the $SU(2)$ case,  for all  $N\geq 3$, the neutral bions 
lead to a center-stabilizing potential.  For $N\geq 3$ we found a first 
order phase transition consistent with lattice gauge theory results. 

Finally, we note that the dependence of $T_c$ on the topological angle $\theta$ can also be studied within our approach, resulting in a decrease of $T_c$,    in agreement with recent lattice studies \cite{D'Elia:2012vv}. 
Details of the calculations  including $\theta$-dependence and higher rank groups will be given elsewhere.

\acknowledgments
We thank Philip Argyres and  Larry Yaffe for useful discussions and Jo\~ ao Penedones for correspondence on the manuscript. 
E.P. thanks SFSU, KIAS, and APCTP for hospitality during the completion 
of this paper and is indebted to Piljin Yi and Yoon Pyo Hong  for many 
useful discussions. We thank the referee for pointing out the 
non-cancellation of the nonzero mode determinants discussed in
the appendix. This work was supported in part by the US Department 
of Energy grant DE-FG02-03ER41260 (T.S.) and the National Science 
and Engineering Research Council of Canada (E.P.).

\appendix 
\section{Supersymmetry and the non-cancelling nonzero-mode determinants}
\label{nonzerodets}

 In this work, we have derived the effective lagrangian for the Polyakov line 
and the dual photon by expanding around the supersymmetric limit of a massless 
Weyl fermion. In deriving equ.~(\ref{bosonic},\ref{monopolepotential}), 
we have used the cancellation of nonzero mode determinants in the 
$N=1$ supersymmetric theory. This cancellation has also been used
in many other instances in the literature, for example in the calculation
of the exact superpotential and gluino condensate in $N=1$ supersymmetric
gauge theory on $\R^3 \times \S^1$ \cite{Davies:1999uw,Davies:2000nw}.

 In this Appendix we discuss an important issue that has not received 
proper attention within the context of $N=1$ supersymmetric theories 
on $\R^3 \times \S^1$ --- the fact that the nonzero mode determinants
in the field of a monopole do not precisely cancel. While we will 
argue that this phenomenon has no significant effect on our main 
result,  we include this discussion for completeness, as it fills 
a gap in the literature on  $N=1$ theories on $\R^3 \times \S^1$.

 The point is that  in the supersymmetric theory, the determinants of 
nonzero mode fluctuations around the BPS or KK monopole-instantons on 
$\R^3 \times \S^1$ do not  precisely cancel, despite the fact that the 
solutions preserve one-half of the supersymmetry. This noncancellation 
occurs  essentially because of the slow fall-off of the monopole-instanton 
background at infinity in $\R^3$ \cite{Weinberg:1979ma, Kaul:1984bp}. 
The nonvanishing of the nonzero mode determinants around supersymmetric 
monopole-instanton backgrounds in $N=1$ theories on $\R^3 \times \S^1$ 
is the $N=1$ counterpart of the mass and central-charge 
renormalization\footnote{See \cite{Rebhan} for a recent review and 
references.}  of BPS monopoles in 4d $N=2$ supersymmetric theories. 
As we will see, the non-cancellation of the nonzero mode determinants 
is  perfectly consistent and is, in fact, required by supersymmetry and 
holomorphy. 

 In the context of purely 3d theories with extended supersymmetry, obtained 
by reducing $N=2$ 4d theories, this non-cancellation has been known since
 \cite{Dorey:1997ij}. However, it was not  addressed in the original 
calculation of nonperturbative effects in  $N=1$ theories on $\R^3 
\times \S^1$  \cite{Davies:1999uw,Davies:2000nw}.
 Only recently, a relevant  calculation 
on $\R^3 \times \S^1$ appeared in the literature \cite{Chen:2010yr}, in 
the context of  theories with $N=2$ supersymmetry. While $N=2$ theories  
share many features with the $N=1$ theory of interest to us,  the most 
important difference is that the branch of moduli space considered in 
\cite{Chen:2010yr} is the 4d Coulomb branch, where the abelianization 
of the gauge group is due to a nonzero expectation value $|a|$ of the  
adjoint Higgs matter supermultiplet. In particular, $|a| \gg \Lambda_{N=2}$ 
was required for consistency of the calculation of \cite{Chen:2010yr} (while  an arbitrary $|a| L$ was allowed). In contrast,  we are interested 
in the $N=1$ theory  where the adjoint Higgs field is absent and the 
abelianization is due, instead,  to a Wilson line expectation value. 
Thus, the result of \cite{Chen:2010yr} is not directly applicable to 
the case of interest to us. 

 From the discussion in the previous paragraph, it is clear that a 
calculation relevant for $N=1$ theories on $\R^3 \times \S^1$ is not 
present in the literature. In what follows, we shall perform this 
calculation. The novel ingredient  that we will use is the  Nye-Singer 
index theorem on $\R^3 \times \S^1$  \cite{Nye:2000eg} in backgrounds 
with nontrivial holonomy, in the form studied by two of us (EP and M\"U) 
in \cite{Poppitz:2008hr}. We will also benefit from insight  gained from 
refs.~\cite{Dorey:1997ij, HoriEtAl,Chen:2010yr}.

 The main object of interest  turns out to be  the ``index" $I(M^2)$, 
defined\footnote{We use quotation marks, since the value of the function 
$I(0)$ is, indeed, the index of the adjoint Dirac operator in a 
monopole-instanton background, but the quantity $I(M^2)$ itself 
depends on $M^2$  \cite{Weinberg:1979ma}.} as:
\begin{equation}
\label{index1}
I(M^2) = \tr {M^2 \over \Delta_- + M^2} - \tr {M^2 \over \Delta_+ + M^2}~,
\end{equation}
where $\Delta_- = D^\dagger D = - D_\mu D^\mu - {1 \over 2} \sigma_{\mu\nu} 
F^{\mu\nu}$ and $\Delta_+ = D D^\dagger = - D_\mu D^\mu $, where $D$ is 
the Weyl operator and $D_\mu$---the covariant derivative in the 
monopole-instanton background. In a self-dual monopole-instanton 
self-dual background, it is the operator $\Delta_-$ that has zero 
modes; see, e.g., \cite{Vandoren:2008xg} for  a review of this notation.
To establish a relation between $I(M^2)$ and the nonzero mode 
determinants, we note the identity:
\begin{equation}
\label{index2}
\int\limits_{\mu^2}^{\Lambda_{PV}^2} {dM^2 \over M^2}\; 
I(M^2) = \tr \ln {\Delta_+ + \mu^2 \over \Delta_+ + \Lambda_{PV}^2} 
-\tr \ln{\Delta_- + \mu^2 \over \Delta_- + \Lambda_{PV}^2}  
= \ln\det {\Delta_+ + \mu^2 \over \Delta_+ + \Lambda_{PV}^2} 
  {\Delta_- + \mu^2 \over \Delta_- + \Lambda_{PV}^2} ~,
\end{equation}
where $\Lambda_{PV}$ is the Pauli-Villars mass and $\mu$ is an auxiliary 
parameter which will be eventually taken to zero. We can now use 
(\ref{index2}) to define:
\begin{equation}
\label{dets1}
R =\lim\limits_{\mu \rightarrow 0} \left(\mu^4 \; e^{\int\limits_{\mu^2}^{\Lambda_{PV}^2} {dM^2 \over M^2}\; I(M^2)} \right)^{3\over 4} = 
\left( { \det \Delta_+ \over \det^\prime \Delta_- } \;  {\det \Delta_- + \Lambda_{PV}^2 \over \det \Delta_+ + \Lambda_{PV}^2 } \right)^{3\over 4}~.
\end{equation}
Here, ``$\det^\prime$" is the determinant with the zero modes omitted,
 i.e., $\det^\prime \Delta_- \equiv \lim\limits_{\mu \rightarrow 0} 
{\det \Delta_- + \mu^2 \over \mu^4}$, using the fact that $\Delta_-$ 
has two zero modes. The quantity $R$ of (\ref{dets1}) is, as is made 
clear from the second identity, equal to the ratio of nonzero modes' 
determinants around monopole-instanton backgrounds appearing in the 
$N=1$ theory. Notice also that $R$  is equal to  the fluctuation 
determinant \cite{Chen:2010yr} for the $N=2$ theory to the power 
of $3/2$. The  various contributions making up the ratio in (\ref{dets1}) 
are as follows: the adjoint fermion contribution is $(\det^\prime 
\Delta_- \det \Delta_+)^{1\over 4}$, while, in a 4d background 
Lorentz gauge, the gauge field determinant  is $(\det^\prime 
\Delta_-)^{-1}$ and the ghost determinant---$(\det \Delta_+)^{1\over 2}$, 
see  \cite{Vandoren:2008xg} for details.

Our main goal here is to compute the ratio of determinants $R$ (\ref{dets1}). We will use the expression for $I(M^2)$ from \cite{Poppitz:2008hr}: 
\begin{eqnarray}
\label{index3}
I(M^2) &=&   I_1 + I_2(M^2) = {2 L v \over \pi} + \sum\limits_{p = - \infty}^\infty \left( {{2 \pi p \over L } + v \over \vert ({2 \pi p \over L } + v)^2 + M^2\vert^{1\over 2}} - {{2 \pi p \over L } -  v \over \vert ({2 \pi p \over L } - v)^2 + M^2\vert^{1\over 2}}\right)~,\nonumber \\
I_1 &\equiv& {2 L v \over\pi}~.
\end{eqnarray}
The definition of $I_2(M^2)$ is evident; see also (\ref{index33}) below. The relation between $v$, the expectation value of $A_4$, and $\Delta\theta$, the angular distance between the eigenvalues of the Polyakov loop, is:
\begin{equation}
\label{vdeltatheta}
\Delta\theta =    L v \in (0, 2 \pi)~. 
\end{equation}
We note that ${2 L v\over \pi} =  4 Q$, where $Q$ is the topological charge of the solution, which  equals $1/2$ at the center symmetric point $v = {\pi\over L}$.

If $L$ is set to zero and the $p$-sums restricted to $p=0$, this is exactly the quantity $I(M^2)_{\R^3}$, found in \cite{Weinberg:1979ma}, and  yielding the well-known result $I(0) = 2$ for the adjoint representation. Note that the function $I(M^2)_{\R^3}$ played a role in both \cite{Dorey:1997ij}, where it was used directly, and \cite{Chen:2010yr}, where the related difference between the density of states of $\Delta_-$ and $\Delta_+$, see \cite{Kaul:1984bp}, was used. 
The novelty on $\R^3\times \S^1$ is that a sum over Kaluza-Klein modes and a contribution of the topological charge to the index (the  non integer $I_1$  term in (\ref{index3})) appear.
The first term in (\ref{index3}) is the bulk contribution to the index and the second---the surface term. The surface term is given by 
the  Kaluza-Klein sums in (\ref{index3}), defined using  zeta-function regularization; in fact, the second term in (\ref{index3}), 
\begin{equation}
\label{index33}
I_2(M^2)= \sum\limits_{p = - \infty}^\infty \left( {{2 \pi p \over L } + v \over \vert ({2 \pi p \over L } + v)^{2}  + M^2\vert^{1\over 2}} - {{2 \pi p \over L } -  v \over \vert ({2 \pi p \over L } - v)^{2} + M^2\vert^{1\over 2}}\right) ~,
\end{equation}
as shown in detail in \cite{Poppitz:2008hr}, leads to:
\begin{eqnarray}
\label{index34}
I_2(0) &=&  \sum\limits_{p = - \infty}^\infty \left( {\rm sign} \left({2 \pi p \over L } + v\right)  - (v \rightarrow - v) \right) =
   - {2 L v \over \pi} - 2 \lfloor - {v L \over 2 \pi} \rfloor+ 2 \lfloor {v L \over 2 \pi} \rfloor ,
\end{eqnarray}
where $ \lfloor x \rfloor$ is the largest integer smaller than $x$. Thus, the topological charge contribution to the index, $I_1$, is canceled by the non integer contribution from the KK sum, $I_2(0)$, and the index equals $2$ for $0 < vL < 2 \pi$. 

\subsection{Calculating the ratio of nonzero mode determinants}

 In the previous Section, we showed that $I_2(0)=2$. Here, we 
  use the expression for $I_2(M^2)$ to compute the ratio of non-zero mode determinants $R$.
We begin with the expression   for $R$ given in (\ref{dets1}), multiplied by 
$e^{- S_0} = e^{ - {4 \pi v L \over g^2(\Lambda_{PV})}}$, where $S_0$ 
is the classical action of a BPS monopole:
 \begin{eqnarray}
 \label{index32}
 e^{- S_0} R &=&  e^{- S_0} \lim\limits_{\mu \rightarrow 0} \left( \mu^4 e^{I_1 \log {\Lambda_{PV}^2 \over \mu^2 }} e^{ \int\limits_{\mu^2}^{\Lambda_{PV}^2} {I_2 (M^2) \over M^2} dM^2 }\right)^{3 \over 4}   \nonumber
 \end{eqnarray}
 \begin{eqnarray}
 &=& e^{- {4 \pi v L \over g^2(\Lambda_{PV})}+  {3 L v \over \pi} \log{\Lambda_{PV}  L}} \lim\limits_{\mu \rightarrow 0} \left( \mu^4 e^{ - I_1\log \mu^2 L^2+  \int\limits_{\mu^2}^{\Lambda_{PV}^2} {I_2 (M^2) \over M^2} dM^2 } \right)^{3 \over 4} 
 \end{eqnarray}
 \begin{eqnarray}
 &=& e^{-  {v L\over \pi} {4 \pi^2 \over g^2(1/L)} } \lim\limits_{\mu \rightarrow 0} \left( \mu^4 e^{ - I_1\log \mu^2 L^2+  \int\limits_{\mu^2}^{\Lambda_{PV}^2} {I_2 (M^2) \over M^2} dM^2 } \right)^{3 \over 4} \equiv  e^{-  {v L\over \pi} {4 \pi^2 \over g^2(1/L)} } \; R_2 ~,\nonumber
 \end{eqnarray}
 where $R_2$ is implicitly defined in the last line above.
  In other words, the UV divergent contribution to $R$ serves the purpose to renormalize the coupling from $\Lambda_{PV}$ to the scale $1\over L$. Any other UV divergence at one loop would be in need of a counterterm and there is not another one at one loop.
    
  Next, we consider $\log R_2$ (omitting the explicit mention of the $\mu \rightarrow 0$ limit to be taken at the end and noticing that the upper limit of the integral can be taken to infinity):
   \begin{eqnarray}
   \label{logr20}    \log R_2 = \log \mu^3 - {3 I_1 \over 2} \log \mu L +{3 \over 2} \int\limits_{\mu}^\infty {dM \over M} \left(\sum\limits_p { {2 \pi p \over L} + v \over 
   | ({2 \pi p \over L} + v)^{2} + M^2|^{1\over 2}} - (v \rightarrow - v)\right) .
   \end{eqnarray}
   Next,  noting that $\int_\mu^\infty {d M \over M \sqrt{A^2 + M^2}} = {1 \over |A|} {\rm arcsinh} { |A| \over \mu} = {1\over |A|} \ln {2 |A| \over \mu} + {\cal{O}}(\mu^2)$, we find:
   \begin{eqnarray}
   \label{logr2}
    \log R_2&=& \log \mu^3 - {3 I_1 \over 2} \log \mu L +{3 \over 2} \sum\limits_p  \left(  {\rm sign}({2 \pi p \over L} + v) \log{ 2 \; |{2 \pi p \over L} + v|  \over \mu} 
 - (v \rightarrow - v)    \right) \nonumber~\\
&=& \log \mu^3 - {3 I_1 \over 2} \log \mu L + {3 \over 2}  \left[ \sum_p {{\rm sign} (p + {v L \over 2 \pi})} - (v \rightarrow - v) \right] \log {4 \pi \over \mu L}    \\
&& ~~~~~~~~~~ ~~~~~~~~~~ ~~~~~+ {3 \over 2}  \left[ \sum_p {{\rm sign} (p + {v L \over 2 \pi}) } \log \vert p + {v L \over 2 \pi}\vert - (v \rightarrow -v ) \right]~. \nonumber 
 \end{eqnarray}
 Before continuing, we recognize from (\ref{index34}) that one of the KK sums appearing in (\ref{logr2}), $\sum_p {{\rm sign} (p + {v L \over 2 \pi})} - (v \rightarrow - v)= 2 - I_1$, giving rise, after substitution in (\ref{logr20}), and remembering $I_1 = {2Lv\over \pi}$:
 \begin{eqnarray}
 \label{logr201}
& \log R_2 & \nonumber \\
&=& \log \mu^3 - {3 I_1 \over 2} \log \mu L + {3 \over 2}  (2 - I_1 ) \log {4 \pi \over \mu L}   + {3 \over 2}  \left[ \sum_p {{\rm sign} (p + {v L \over 2 \pi}) } \log \vert p + {v L \over 2 \pi}\vert - (v \rightarrow -v ) \right]~  \nonumber \\
&=&3 \log {4 \pi \over L} - {3} {L v \over \pi}  \log 4 \pi + {3 \over 2}  \left[ \sum_p {{\rm sign} (p + {v L \over 2 \pi}) } \log \vert p + {v L \over 2 \pi}\vert - (v \rightarrow -v ) \right]~.
 \end{eqnarray}
 Now we can deal with the remaining KK sum, by  writing it as an $s$-derivative of a function, evaluated at $s=0$: 
 \begin{eqnarray}
 \label{logr202} & \log R_2 & \nonumber \\
&=&  3\log {4 \pi \over L} - {3} {L v \over \pi}  \log 4 \pi  - {3 \over 2}   {d \over d s}\left[ \sum_p { {\rm sign} (p + {v L \over 2 \pi}) \over \vert p + {v L \over 2 \pi}\vert^{s}} - (v \rightarrow -v ) \right]\bigg\vert_{s \rightarrow 0}~.
 \end{eqnarray}
Next, we define the function: 
 \begin{equation}
 \label{hfunction}
  H(s, a) =  \sum_p  {{\rm sign}(p + a) \over |p + a|^s}~,
 \end{equation}
 and rewrite (\ref{logr202}) as: 
 \begin{equation}
 \label{logr203} 
  \log R_2 = 3\log {4 \pi \over L} - {3} {L v \over \pi}  \log 4 \pi  - {3 \over 2}  \lim\limits_{s \rightarrow 0} \;  {d \over d s} \left( H(s, {v L \over 2 \pi})   - H(s, -{v L \over 2 \pi}) \right)~.
  \end{equation}
 Now,   for $1 > a > 0$ we have:
  \begin{equation}
  \label{hfn1}
  H(s, a) = \sum_{p \ge 0} {1 \over |p + a|^s} - \sum_{p \ge 0} {1 \over |p + 1 - a|^s} =    \zeta(s, a) - \zeta(s,  1- a) ~,
  \end{equation}
where $\zeta(s,a) = \sum_{p\ge 0} |p + a|^{-s}$ is the incomplete zeta function, and similar for:
    \begin{equation}
  \label{hfn2}
  H(s, - a) = \sum_{p \ge 0} {1 \over |p + 1- a|^s} - \sum_{p \ge 0} {1 \over |p+a|^s} =    \zeta(s, 1 -a) - \zeta(s, a) ~.
    \end{equation}
Thus, $R_2$ is given by:
   \begin{equation}
  \label{logr22}
  \log R_2 =   3\log {4 \pi \over L} - {3} {L v \over \pi}  \log 4 \pi   + 3 \left(\zeta^\prime (0, 1 - {Lv\over 2 \pi}) - \zeta^\prime(0, {Lv\over 2 \pi})\right)~,  
  \end{equation}
which, upon plugging into (\ref{index32}), and using  $\zeta^\prime (0,x) = \ln \Gamma(x) - {1 \over 2} \log 2 \pi$ yields for $e^{- S_0}$ times the ratio of determinants $R$ in the BPS monopole background:
  \begin{eqnarray}
  e^{-S_0} R\bigg\vert_{BPS} &=& \left( {4 \pi \over L}\right)^3 e^{-  {v L\over \pi} \left[{4 \pi^2 \over g^2(1/L)} + 3 \log 4 \pi\right]+ 3   \left(\zeta^\prime (0, 1 - {Lv\over 2 \pi}) - \zeta^\prime(0, {Lv\over 2 \pi})\right)} \nonumber \\
  &=&   \left( {4 \pi \over L}\right)^3  e^{-  {v L\over \pi} {4 \pi^2 \over g^2({4 \pi\over L})} + 3 \log\Gamma\left(1 - {L v \over 2 \pi}\right)- 3 \log\Gamma \left({v L\over 2 \pi}\right) }~. \label{rr22}
  \end{eqnarray}
  For the ratio of determinants around the KK monopole, we replace ${vL\over 2\pi} \rightarrow 1 - {vL\over 2\pi}$, 
  yielding:
  \begin{eqnarray}
  \label{rr22KK}
  e^{-S_0} R\bigg\vert_{KK}  =   \left( {4 \pi \over L}\right)^3  e^{- {8 \pi^2 \over g^2({4 \pi \over L})} +  {v L\over \pi} {4 \pi^2 \over g^2({4 \pi\over L})} - 3 \log\Gamma\left(1 - {L v \over 2 \pi}\right)+ 3 \log\Gamma \left({v L\over 2 \pi}\right) }~.
  \end{eqnarray}

 We can now study the expansion of the $\Gamma$ functions from (\ref{rr22}) and (\ref{rr22KK}) near the center symmetric points:
   \begin{eqnarray}
   \label{gammas}
   3 \log{ \Gamma\left(1 - {L v \over 2 \pi}\right) \over \Gamma \left({L v \over 2 \pi}\right) }&=& -  3 \psi(1/2) \left( {L v \over \pi} -1 \right)  + {\cal{O}}(({L v \over \pi} - 1)^3)  \nonumber \\ &\approx&  5.8903 \left( {L v \over \pi} -1 \right) ,
   \end{eqnarray} 
  where we used the value $\psi(1/2)= \Gamma^\prime(1/2)/\Gamma(1/2) \approx -1.96É$ of the digamma function. Thus, we can rewrite (\ref{rr22}) as:
 \begin{eqnarray}
 \label{rr231}
  e^{-S_0} R\bigg\vert_{BPS} =  \left( {4 \pi \over  L}\right)^3  e^{- {4 \pi^2 \over g^2({4 \pi\over L})} - ({v L \over \pi} - 1) \left( {4 \pi^2 \over g^2({4 \pi\over L})} + 3 (\log\Gamma(1/2))^\prime \right) + {\cal{O}}(({L v \over \pi} - 1)^3) }~,
 \end{eqnarray}
 as well as a similar expression for the KK monopole (\ref{rr22KK}).
   Thus, near the center symmetric point, we have that there is a small shift of the scale of the coupling constant away from $4 \pi/L$ (using $-1.96 \approx - \log 7.12$, to a scale a few times lower than $1/R= 2\pi/L$) and that 
the exponential in    $e^{- S_0} R$ is a linear function of the  deviation from the center symmetric vacuum, $ {v L \over \pi} - 1$, up to cubic terms. The behavior of the ratio of one-loop determinants is illustrated for general $vL$ in Figure~\ref{fig:phase4}.

    \begin{FIGURE}[h]{
    \parbox[c]{\textwidth}
        {
        \begin{center}
        \includegraphics[angle=0, scale=1.3]{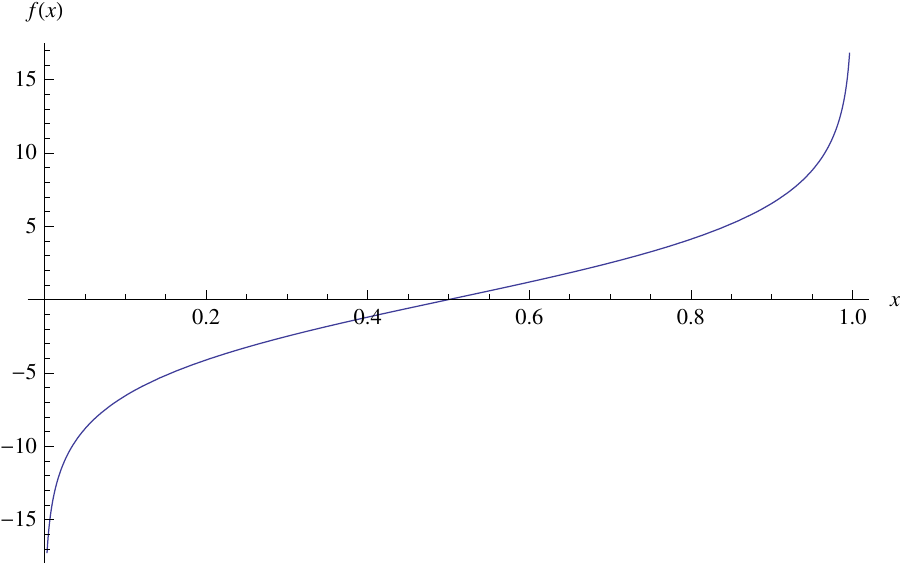}
	\hfil
        \caption
	{The behavior of the ratio of nonzero mode determinants around a BPS monopole-instanton, eqn.~(\ref{rr22}). We define $f(x) = 3(\log \Gamma(1-x) - \log \Gamma(x))$, where $x = vL/2\pi$. The singularities  of $f(x)$ near the gauge-symmetry restoration points ($x=0,1$) are logarithmic and are, in fact, required in order to take the proper 3d limit, see discussion around eqn.~(\ref{rr23})). Notice that at the center-symmetric point the nonzero modes determinants cancel exactly, $f(1/2)=0$, while the slope there is $-3 \psi(1/2) \simeq 5.89É$, see eqn.~(\ref{gammas}).} 
        \label{fig:phase4}
        \end{center}
        }
    }
\end{FIGURE}

It is of interest to also consider the neighborhood of $vL  = 0$, in particular, if we wish to take the 3d limit and compare with previous studies as a check on our calculation. In the 3d limit, we take $L \rightarrow 0$, but keep $g_3^2 = L/g_4^2$ fixed; as before, $v$ is an arbitrary position on the moduli space, which  becomes noncompact in the 3d limit. The contribution of the KK monopole vanishes in the 3d limit, as its action becomes infinite. Hence, we concentrate on the BPS monopole amplitude. In this limit, we have that:
      \begin{eqnarray}
   \label{gammas1}
   3 \log{ \Gamma\left(1 - {L v \over 2 \pi}\right) \over \Gamma \left({L v \over 2 \pi}\right) }\bigg\vert_{Lv \approx 0}&=& -  3 \log { 2 \pi \over L v}  + 3 \gamma_E {L v \over \pi} +  {\cal{O}}(({L v \over \pi} )^3)  ,
   \end{eqnarray} 
   so we can write for (\ref{rr22}):
   \begin{eqnarray}
 \label{rr23}
  e^{-S_0} R\bigg\vert_{BPS} &=&  \left( {4 \pi \over L}\right)^3  e^{- {v L \over \pi}{4 \pi^2 \over g^2({4 \pi \over L})}  - 3 \log {2 \pi \over v L} + 
  3 \gamma_E {L v \over \pi} +  {\cal{O}}(({L v \over \pi} )^3) }~\nonumber \\
  &\rightarrow &\left( {2 v  } \right)^3 e^{ - {4 \pi v \over g_3^2}}~, ~~ {\rm with} ~ ~ L\rightarrow 0, ~~ {\rm fixed} ~ g_3^2 = {L \over g_4^2({4\pi \over L})}~.
 \end{eqnarray}
We note that in this limit, our calculation exactly reproduces the 3d result of \cite{Dorey:1997ij}. To facilitate the comparison, we need to  take into account the facts that $v=m_W$ in our normalization and that our calculation is in the 4-supercharges theory, not the 8-supercharges one,
 which changes the overall power of the dimensional pre factor from 2, as in eqn.~(26) from \cite{Dorey:1997ij}, to 3 as in our (\ref{rr23}). 
  It is clear from the above that the noncancellation of the determinants is required in order to have a smooth 3d limit.
    
\subsection{Interpretation} 

 The main question is, what is the interpretation of the cubic and higher terms  in $f(x)$ away from the center symmetric point? It appears at first sight that  they do not belong in the superpotential, as they are not holomorphic functions of $vL + i \sigma$. Further, any nonlinear terms in this variable appearing in the exponent would be in conflict with symmetry arguments.  
 
We note that
these questions arise already in the 3d supersymmetric theories and have been addressed in some cases. Notably,  
in the 8-supercharges 3d theory, the BPS monopole-instanton induced terms give rise to the four-fermi interaction \cite{Dorey:1997ij}:
 $$
 {v \over (g_3^2)^4}\; \lambda^2 \psi^2 e^{ - {4 \pi v \over g_3^2} + i \sigma}~.
 $$
 Since the 8-supercharge theory admits no superpotential, this term, including the 
 $v$-dependent prefactor should be interpreted as arising from a component expansion of a supersymmetric sigma model with a hyper-K\" abler metric. In fact, as shown in \cite{Dorey:1997ij}, the  above four-fermi term is fully consistent with the semiclassical expansion of the  Atiyah-Hitchin metric.  
 
 On the other hand,
 in the 4-supercharges 3d theory, we have, from (\ref{rr23}), after incorporating bosonic and fermionic zero modes, the following form for the   monopole-instanton induced fermion vertex:
 $$ 
{ v^3 \over (g_3^2)^3} \; \lambda^2 e^{ - {4 \pi v \over g_3^2} + i \sigma}~.
 $$
 It is natural to ask how this term is absorbed in an  $e^{- X}$ superpotential, where $X = {4 \pi v \over g_3^2} - i \sigma$, as argued long time ago \cite{Affleck:1982as}. The pre-exponential factor there and in subsequent work was assumed to be a modulus-independent constant  and has not been calculated before, to the best of our knowledge. Thus the question regarding the incorporation of the nonholomorphic $\sim v^3$ prefactor in a supersymmetric effective lagrangian applies to the 3d 4-supercharge theory as well.  
 
 We now proceed to address this issue by first writing down the fermion vertices that accompany the BPS (\ref{rr22}) and  KK (\ref{rr22KK}) monopole-instantons, performing the same steps as in  \cite{Davies:2000nw}. We skip the details (of  including the collective coordinates measure, etc., as they are identical to \cite{Davies:2000nw}) and note that the only difference is that we replace the Pauli-Villars scale appearing in the fermion measure there with the ratio of determinants that we calculated above. The fermion bilinear terms---the kinetic term and the ones that arise due to the  BPS and KK monopole-instantons---are (all couplings below are taken at the scale $4 \pi\over L$):
 \begin{eqnarray}
 \label{fermbilinear1}
 L_{ferm} &=& {L \over g^2} i \bar\lambda \sigma^m \partial_m \lambda  \\
 && +\left({L \over g^2}\right)^2 {2^4 \pi^2 L \over g^2} \left({4 \pi \over L}\right)^3\left[ e^{- {v L \over \pi} {4 \pi^2 \over g^2} + i \sigma + f({v L \over 2 \pi}) } +   e^{- {8 \pi^2 \over g^2} +  {v L \over \pi} {4 \pi^2 \over g^2} - i \sigma - f({v L \over 2 \pi})} \right] \lambda \lambda ~\nonumber~,
 \end{eqnarray}
 where 
 \begin{equation}
 \label{ef}
e^{f(x)} = \left(\Gamma(1-x)\over \Gamma(x)\right)^3
 \end{equation}
 was already defined in the caption of Figure \ref{fig:phase4}. For future interpretation, it is also useful to recall the original, not resummed, expression for $f(x)$. This follows from (\ref{logr201}) and, as given below, is valid for $0<x<1$:
 \begin{equation}
 \label{ef2}
 f(x) = 3 \ln x + 3 \sum\limits_{p>0} \ln(p+x) - \ln(p - x) ~.
 \end{equation}
 Next, we rewrite (\ref{fermbilinear1}), using $b = {4 \pi \over g^2} v L$ and expanding around the center-symmetric value
 $b \equiv {4 \pi^2 \over g^2} + b^\prime$ (in other words, $b^\prime$ is the fluctuation around the center symmetric vev):
  \begin{eqnarray}
 \label{fermbilinear2}
 L_{ferm} &=& {L \over g^2} i \bar\lambda \sigma^m \partial_m \lambda + {1 \over 2} {g^2 \over 16 \pi^2 L}\; \partial_m b^\prime \partial^m b^\prime + \ldots\\
 && + {2^{10} \pi^5  \over g^6}~e^{- {4 \pi^2 \over g^2}} \left[ e^{- b^\prime + i \sigma} e^{f\left({1 \over 2 } + {g^2 \over 8 \pi^2} b^\prime\right) } +   e^{b^\prime - i \sigma} e^{ - f\left({1 \over 2 } + {g^2 \over 8 \pi^2} b^\prime\right) } \right]\lambda \lambda ~\nonumber~,
 \end{eqnarray}
 where we included the kinetic term for $b^\prime$ to leading order (subleading contributions to the $b^\prime$ kinetic term, to be elaborated below, are denoted by dots).
 The main question we want to address is how (\ref{fermbilinear2}) can be incorporated in a supersymmetric effective lagrangian, given the extra, seemingly nonholomorphic, dependence  of the fermion 't Hooft vertex on $b^\prime - i \sigma$, arising from the noncanceling determinants around the BPS and KK monopole-instantons. 
 
 For the reader not so interested in the details, 
 the summary of the discussion of the following section is that after properly performing the supersymmetric ``photon-dual photon" duality transformation, the dimensionless chiral superfield $B$ dual to the abelian vector multiplet (in 3d, equivalent to a real linear multiplet), describing $A_4$ and the 3d photon, is such that its lowest component is:
 \begin{equation}
 \label{x1}
 B\vert = b^\prime - f\left({1 \over 2 } + {g^2 \over 8 \pi^2} b^\prime\right)  - i \sigma~.
 \end{equation}
 In terms of the chiral superfield (\ref{x1}), the superpotential is holomorphic and is of the well-known affine-Toda form, $W \sim e^{-B } + e^{B}$. In the next Section, we discuss the details of the duality transformation and the interpretation of the function $f$ in (\ref{x1}) as encoding perturbative corrections to the moduli space metric on $\R^3 \times \S^1$. We recall again that $f$ is nonzero only away from the center-symmetric point, thus the extremal point of  $W(B)$, $B=0$,  still corresponds to the center-symmetry preserving vacuum $b^\prime = 0$, in accord with the conclusions from the earlier work \cite{Davies:1999uw,Davies:2000nw}.

 \subsection{Linear-chiral superfield duality}
 
 The dimensional reduction of an abelian 4d vector multiplet to 3d is described as a real linear multiplet, $W$, defined as:\footnote{Our notation, including the  supercovariant derivatives and $V$, is that of Wess and Bagger \cite{WB}.}
 \begin{equation}
 \label{w1}
 W =  {1 \over 2} \;  \bar\sigma^{3\;  \dot\alpha \alpha} \bar{D}_{\dot\alpha} D_\alpha V ,
 \end{equation}
 where $V$ is the usual real vector superfield in 4d (dimensional reduction in (\ref{w1}) as written is performed along the $x^3$ direction and, hence, the lowest component of (\ref{w1}) is $W| = A_3$). The field $W$ obeys the linear multiplet relations $D^2 W = \bar{D}^2 W = 0$ and is  invariant under the usual (3d) supergauge transformations $V \rightarrow V - i (\Lambda - \Lambda^\dagger)$. The lowest component of $W$, as already indicated, is the real scalar in the 3d vector mulitiplet (i.e., is related by rescaling to our fields $b$ or $b^\prime$), while the other terms in its superspace expansion involve the fermions and the $U(1)$ gauge field strength. 
 As defined above, the dimension of $W$ is unity. The minimal kinetic term for the vector multiplet is given by the first term in the  $D$-term action given below:
 \begin{equation}
 \label{w2}
 \int d^4 \theta d^3 x \left( - {1 \over 2 e^2} W^2 - F(W) \right)~,
 \end{equation}
 while the function $F(W)$ incorporates possible nonlinear corrections to the kinetic terms. 
 The action (\ref{w2}) is the most general two-derivative one involving the linear multiplet (note that $W$ can not appear in integrals over half-superspace). The coupling $e^2$  in (\ref{w2}) has unit mass dimension  and denotes the 3d gauge coupling; clearly, $e^2$ or another  mass scale must be present in $F(W)$.
The kinetic term of the lowest component of $W$, which we denote by $w$ in this Section ($W\vert \equiv w$), hoping not to cause confusion, reads:
 \begin{equation}
 \label{w21}
  \int d^3 x\;  {1 \over 2} \left( {1 \over e^2} + F^{\prime\prime}(w) \right)  \partial_m w \partial^m w ~.
 \end{equation} 
 For the theory to make sense, the function ${1 \over e^2} + F^{\prime\prime}(w)$ (the ``tau-parameter," or moduli space metric), which also determines the effective $U(1)$ coupling along the Coulomb branch, has to be positive.
 
  Linear-chiral duality is performed via  a Legendre transformation as follows:
 \begin{equation}
 \label{w3}
 \int d^4 \theta d^3 x \left( -{1 \over 2 e^2} W^2 - F(W) + (B^\dagger + B) W \right)~.
 \end{equation}
The  dimensionless chiral superfield $B$ is introduced as a  Lagrange multiplier and $W$ is now regarded as an unconstrained real superfield. Variation with respect to $B$ and $B^\dagger$, taking into account the chirality constraint $D_\alpha B^\dagger = \bar{D}_{\dot\alpha} B = 0$,  enforces the linear multiplet conditions $\bar{D}^2 W = D^2 W = 0$ on the real superfield $W$. 

On the other hand, a variation of (\ref{w3}) with respect to $W$ gives:\footnote{Notice that this relation also implies that the fermion components of $B$ are slightly different from the fermionic components of $W$; we will ignore this in what follows.}
\begin{equation}
\label{w4}
B + B^\dagger = {1 \over e^2} W + F^\prime(W)~.
\end{equation}
The equation of motion (\ref{w4}) can be inverted, e.g. perturbatively, to give $W  = e^2 (B + B^\dagger) + \ldots$, and upon plugging the solution of (\ref{w4}) in (\ref{w3}), one is left with a dual theory of chiral superfields given by:
 \begin{equation}
 \label{w5}
 \int d^4 \theta d^3 x \; K(B + B^\dagger)~=  \int d^4 \theta d^3 x  \left({e^2 \over 2 }(B + B^\dagger)^2 + \ldots\right)~=  \int d^4 \theta d^3 x  \left({e^2} B^\dagger B + \ldots\right),
 \end{equation}
 where the dots denote higher order terms determined by the form of $F(W)$. For a recent reference, see, for example, \cite{HoriEtAl}.
 The point of the discussion above is that the relation between the real part of the  lowest component of the dual chiral superfield $B$ and the lowest component of the linear superfield $W$ is simple---i.e., linear---only if the function $F(W)$ is quadratic, i.e., if there are no nontrivial corrections to the kinetic terms of the $b^\prime \sim w$ field. In general, this relation can be complicated. 
 
 In particular, in our case of interest,  the duality transformation on $\R^3 \times \S^1$ is performed after Kaluza-Klein modes are integrated out. Integrating out the KK modes generates nontrivial corrections to the kinetic terms of $A_4$ and 3d photon. 
 As we will now see, these perturbative corrections are precisely related to the function $f$ (\ref{ef2}) representing the ratio of nonzero modes' fermionic and bosonic determinants around the BPS and KK monopole instantons.  
To see this, let us now apply the duality transformation to our theory. As already noted, 
 eqn.~(\ref{w4}) determines  the lowest component of $X$ in terms of $w(x)$ and the dual photon $\sigma(x)$:
 \begin{equation}
 \label{x2}
 B\vert = {1 \over 2 e^2} w + {1 \over 2} F^\prime(w) - i \sigma~,
 \end{equation}
which, upon comparison to (\ref{x1}) immediately implies that:
\begin{equation}
\label{x3}
b^\prime(x) \equiv {1 \over 2 e^2} \; w(x)~~ {\rm and} ~ F^\prime(w) \equiv - 2 f\left({1\over 2} + {g^2 \over 16 \pi^2 e^2} w \right)~.
\end{equation} 
Then, we can use (\ref{x3}) to compare the leading order kinetic term of $w$, eqn.~(\ref{w21}), with that of $b^\prime$, eqn.~(\ref{fermbilinear2}), and find:
\begin{equation}
\label{x4}
e^2 = {g^2 \over 64 \pi^2 L}~.
\end{equation}

We can now use the explicit form (\ref{ef2}) of $f$ and (\ref{x4}) to find that the corrections to the kinetic term for $w$ (\ref{w21}) are determined by the effective ``tau parameter":
 \begin{eqnarray}
 \label{tau1}
  {1 \over e^2} + F^{\prime\prime}(w) &=&  {64 \pi^2 L \over g^2}  - 8 L  f^\prime\left({1\over 2} + 4 L w \right) \\
  &=&{64 \pi^2 L \over g^2}  - 24 L \left[{1 \over x} + \sum\limits_{p>0} {1 \over p - x} +
   {1 \over p + x} \right]\bigg\vert_{x \rightarrow {{1 \over 2} + 4 L w}}~.
\end{eqnarray}
 In the last line,  we used eqn.~(\ref{ef2}) for $f(x)$ and denoted $f^\prime(x) \equiv d f(x)/dx$.
 
  In order to interpret the corrections, we rewrite (\ref{tau1}) as:
 \begin{eqnarray}
 \label{tau2}
 {1 \over e^2} + F^{\prime\prime}(w) &=& 64 \pi^2 \left( {L \over g^2} - {3 L\over 8 \pi^2} 
 \left[{1 \over x} + \sum\limits_{p>0} {1 \over p - x} +
   {1 \over p + x} \right]\bigg\vert_{x \rightarrow {{1 \over 2} + 4 L w}}~\right)
 \end{eqnarray}
 It is most straightforward to first consider the purely-3d limit $L \rightarrow 0$, ${L\over g^2}$-fixed. In this case, the sum over $p>0$ in (\ref{tau2}) drops out and only the first term in the square brackets  survives. Recall that $g^2$ is taken at the scale ${4 \pi\over L}$, which is now the 3d UV cutoff scale (the bare 3d coupling is ${L\over g^2}$).
 The coupling (\ref{tau2}) can be rewritten, using $x = {v L \over 2 \pi}$, as:
 \begin{eqnarray}
 \label{tau3}
 {1 \over e^2} + F^{\prime\prime}(w) &=& 64 \pi^2 \left( {L \over g^2} - {3 L \over 8 \pi^2} {2 \pi \over v L }\right)
 = 64 \pi^2 \left( {L \over g^2} - {3 \over 4 \pi  v} \right)~.
 \end{eqnarray}
  The ${3\over4 \pi v}$ shift  of ${L \over g^2}$ in (\ref{tau3}) represents exactly the one-loop shift of the bare 3d gauge coupling ${L\over g^2}$ due to integrating out the 3d heavy vector multiplet---the heavy $W$-boson and superpartner, of mass $\sim v$, along the Coulomb branch parameterized by $v$  in an $SU(2)$ theory. This one-loop shift was calculated in \cite{Smilga:2004zr}, see eqn.~(83) there.\footnote{The factor of $64 \pi^2$ represents an overall normalization factor; note that ref.~\cite{Smilga:2004zr} calculated precisely the ``running" of the 3d coupling from the bare value $L\over g^2$ to the value at scales below $v$, with a result exactly as given in the brackets in (\ref{tau3}).}
    Note that we
   obtained the same result as \cite{Smilga:2004zr} by a rather roundabout way---by calculating the ratio of determinants of nonzero modes in the monopole-instanton background and then demanding consistency of the result with holomorphy of the 
   superpotential. While this may appear miraculous, the matching of the results had to be true, by the power of supersymmetry. 
   
  Going back to finite-$L$, i.e. to $\R^3 \times \S^1$, the interpretation of the other terms appearing in (\ref{tau2}) is now clear. The sum over $p>0$ represents the effect of the Kaluza-Klein partners of the heavy vector multiplets on the unbroken-$U(1)$ gauge coupling; recall that at every KK level with $p>0$, there are two heavy vector multiplets of mass ${2 p \pi \over L} \pm v$. We are not aware of an explicit perturbative calculation of this in a compactified $N=1$ theory \ however, similar to the purely-3d four supercharge case  and to the $N=2$ 4d theory,\footnote{See Section~4.1 in  \cite{Gaiotto:2008cd} for an  expression, similar to (\ref{tau2}), for the one-loop correction to the moduli space metric of the $N=2$ theory on $\R^3 \times \S^1$, valid also on the Coulomb branch with only a Wilson line turned on.} and given the suggestive form of the KK sum in (\ref{tau2}),  it is natural to conjecture that, by supersymmetry, the nonvanishing ratio of nonzero modes determinants around the monopole-instanton solutions precisely encodes the one-loop corrections to the $U(1)$ gauge coupling.

 In our  discussion in the main text, we have ignored the non-canceling nonzero mode determinants. This is because their net effect is to modify the K\" ahler potential for the chiral superfield $B$, see eqn.~(\ref{x2}), by small  loop-suppressed terms. Our main concern is the competition between nonperturbative effects at weak coupling, while the  perturbative deformation of  the K\" ahler potential (\ref{kahler}) provides only small corrections to the leading order semiclassical results.

\section{The monopole-induced potential in softly-broken SYM}
\label{monopolempotential}

 In this appendix, we determine the prefactor of the monopole instanton
amplitude in softly broken $\N=1$ theory on $\R^3\times \S^1$. From 
\cite{Davies:2000nw}, the collective coordinate measure for ${\cal{M}}_1$ 
(BPS) monopole-instantons in $SU(2)$ gauge theory, is:
\begin{equation}
\label{monopolemeasure}
\int d\mu_{BPS} = {M_{PV}^3 L \over 2 \pi g^2} 
    \int d^3 a \; d \phi \; d^2 \zeta \; e^{- b + i \sigma}~,
\end{equation}
where $b = \langle b \rangle + b(a)$, $\sigma = \langle \sigma \rangle 
+ \sigma(a)$. The three-dimensional vector $a$ denotes the collective 
coordinate representing the center of the monopole, $\phi$ is the angular 
$U(1)$ collective coordinate, $\zeta^\alpha$ are the two Grassmann 
fermion zero mode coordinates, and the prefactor represents the product 
of all collective coordinate Jacobians. 

 Now, we add the fermion mass term $\Delta L_m = {m \over g^2}  \tr 
[\lambda \lambda]$, see (\ref{lm}),  to the Lagrangian of SYM. In order 
to obtain the contribution of, for example, the ${\cal{M}}_1$ (or BPS) 
monopole-instanton to the potential for $b$ and $\sigma$, we  saturate 
the integral over the fermion zero modes by a single mass-term insertion:
\begin{eqnarray}
\label{insertion1}
\int d \mu_{BPS}\, e^{ - \Delta S_m} 
 \approx - {M_{PV}^3 L \over g^2} \int d^3 a\, e^{- b + i \sigma } \;  
  {mL  \over g^2} \int d^2 \zeta\, d^3 x \; 
   \tr\left[ \lambda^0(x-a)\lambda^0(x-a)\right]~,
\end{eqnarray}
where  $\Delta S_m=\int d^4x\, \Delta L_m$ and $\lambda^0(x-a)$ is the fermion zero 
mode of the monopole located at $a$. Next we note that  ${L \over g^2} 
\int d^2 \zeta\, d^3 x\, \tr [\lambda^0 \lambda^0]$ is exactly the fermion 
zero-mode Jacobian calculated in \cite{Dorey:1997ij}, which  is equal to 
$2 S_{cl}^{{\cal{M}}_1} = 2 b$. Collecting all factors, we conclude that 
the contribution of a ${\cal{M}}_1$ monopole-instanton to the potential is:
\begin{equation} 
\label{deltaLBPS}
\Delta V^{{\cal{M}}_1} = {2 m M_{PV}^3 L \over g^2} \; b \; 
   e^{- b + i \sigma} ~. 
\end{equation}
Proceeding similarly, we find that the ${\cal{M}}_2$ (or KK) monopole-instanton 
contributes, using its action $S_{cl}^{{\cal{M}}_2} = {8 \pi^2 \over g^2} - b$ 
instead:
\begin{equation}
\label{deltaLKK}
\Delta V^{{\cal{M}}_2} = {2 m M_{PV}^3 L   \over g^2}\; 
  \left({8 \pi^2 \over g^2} - b\right) \; 
  e^{- {8 \pi^2 \over g^2} +  b - i \sigma} ~. 
\end{equation}
To obtain the total ${\cal{O}}(m)$ contribution  to the scalar potential, 
we now sum over the contributions of the two monopole-instantons and their 
complex conjugates:
\begin{eqnarray}
\label{vmon2}
 V_{\rm mon.}  &=& \Delta V^{{\cal{M}}_1} + \Delta V^{{\cal{M}}_2} 
             + {\rm  h.c.}  \\ 
 &=&  {32 \pi^2 m M_{PV}^3 L  \over g^4}  \; e^{ - {4 \pi^2 \over g^2}}\; 
         \cos \sigma \left( \cosh b^\prime  
          - {g^2 \over 4 \pi^2} \; b^\prime \sinh b^\prime\right) ~.
\end{eqnarray}
Finally, we use the relations (\ref{rgrelations}), to cast $V_{\rm mon.}$ 
into the form given in (\ref{monopolepotential}).

\end{document}